\documentclass[twocolumn]{aastex61}
\usepackage{amssymb,amsmath}

\def\apjl{ApJ}

\def\apjs{ApJS}

\def\aap{Astron. and Astrophys.}

\def\xmm{{\it XMM-Newton}}
\def\nustar{{\it NuSTAR}}
\def\nicer{{\it NICER}}
\def\sw{{\it Swift}}
\def\tng{{\it TNG}}
\def\sifap{\it SiFAP}
\def\tngsifap{\it TNG/SiFAP}
\def\xmmepic{\it XMM-Newton/EPIC}
\def\tjo{\it Telescopi Joan Or\'o}
\def\not{\it Nordic Optical Telescope}
\def\swfirst{\it Neil Gehrel's Swift Observatory}
\def\igr{IGR J18245--2452}

\def\1023{PSR J1023+0038}
\def\xss{XSS J12270-4859}
\def\ltsima{$\; \buildrel < \over \sim \;$}
\def\simlt{\lower.5ex\hbox{\ltsima}}
\def\gtsima{$\; \buildrel > \over \sim \;$}
\def\simgt{\lower.5ex\hbox{\gtsima}}
\def\flux{erg~cm$^{-2}$~s$^{-1}$}
\def\arcsec{$^{\prime\prime}$}

\newcommand\aastex{AAS\TeX}

\received{\today}
\submitjournal{ApJ}

\shorttitle{\aastex\ sample article}
\shortauthors{Papitto et al.}

\begin{document}

\title{Pulsating in unison at optical and X-ray energies: simultaneous high-time resolution observations of the transitional millisecond pulsar {\1023}}

\correspondingauthor{A.~Papitto}
\email{alessandro.papitto@inaf.it}
\author[0000-0002-0786-7307]{A.~Papitto}
\affil{INAF--Osservatorio Astronomico di Roma, via Frascati 33, I-00076, Monteporzio Catone (RM), Italy}

\author{F.~Ambrosino}
\affiliation{INAF-IAPS,Via del Fosso del Cavaliere 100, I-00133 Rome, Italy}

\author{L.~Stella}
\affiliation{INAF--Osservatorio Astronomico di Roma, via Frascati 33, I-00076, Monteporzio Catone (RM), Italy}

\author{D.~Torres}
\affiliation{Institute of Space Sciences (ICE, CSIC​)​, Campus UAB, Carrer de Can Magrans, E-08193, Barcelona, Spain}
\affiliation{Institut d'Estudis Espacials de Catalunya (IEEC), E-08034 Barcelona, Spain}
\affiliation{Instituci\'o Catalana de Recerca i Estudis Avan\c{c}ats (ICREA), E-08010 Barcelona, Spain}

\author{F.~Coti Zelati}
\affiliation{Institute of Space Sciences (ICE, CSIC​)​, Campus UAB, Carrer de Can Magrans, E-08193, Barcelona, Spain}

\author{A.~Ghedina}
\affiliation{Fundaci\'{o}n Galileo Galilei - INAF, Rambla Jos\'{e} Ana Fern\'{a}ndez P\'{e}rez, 7, E-38712 Bre\~{n}a Baja, TF, Spain}

\author{F.~Meddi}
\affiliation{Dipartimento di Fisica, Universit\`a di Roma ``La Sapienza'', Piazzale Aldo Moro, 5, I-00185 Roma, Italy}

\author{A.~Sanna}
\affiliation{Dipartimento di Fisica, Universit\`a di Cagliari, SP Monserrato-Sestu, Km 0.7, I-09042 Monserrato, Italy}

\author{P.~Casella}
\affiliation{INAF--Osservatorio Astronomico di Roma, via Frascati 33, I-00076, Monteporzio Catone (RM), Italy}

\author{Y.~Dallilar}
\affiliation{Department of Astronomy, University of Florida, 211 Bryant Space Science Center, Gainesville, FL 32611, USA}

\author{S.~Eikenberry}
\affiliation{Department of Astronomy, University of Florida, 211 Bryant Space Science Center, Gainesville, FL 32611, USA}

\author{G.~L.~Israel}
\affiliation{INAF--Osservatorio Astronomico di Roma, via Frascati 33, I-00076, Monteporzio Catone (RM), Italy}

\author{F.~Onori}
\affiliation{INAF-IAPS,Via del Fosso del Cavaliere 100, I-00133 Rome, Italy}

\author{S.~Piranomonte}
\affiliation{INAF--Osservatorio Astronomico di Roma, via Frascati 33, I-00076, Monteporzio Catone (RM), Italy}

\author{E.~Bozzo}
\affiliation{Department of Astronomy, University of Geneva, Chemin d'Ecogia 16, CH-1290, Versoix, Switzerland}

\author{L.~Burderi}
\affiliation{Dipartimento di Fisica, Universit\`a di Cagliari, SP Monserrato-Sestu, Km 0.7, I-09042 Monserrato, Italy}

\author{S.~Campana}
\affiliation{INAF--Osservatorio Astronomico di Brera, via Bianchi 46, I-23807 Merate (LC), Italy}

\author{D.~de Martino}
\affiliation{INAF--Osservatorio Astronomico di Capodimonte, Salita Moiariello 16, I-80131 Napoli, Italy}

\author{T.~Di Salvo}
\affiliation{Universit\`a degli Studi di Palermo, Dipartimento di Fisica e Chimica, via Archirafi 36, I-90123 Palermo, Italy}

\author{C.~Ferrigno}
\affiliation{Department of Astronomy, University of Geneva, Chemin d'Ecogia 16, CH-1290, Versoix, Switzerland}

\author{N.~Rea}
\affiliation{Institute of Space Sciences (ICE, CSIC​)​, Campus UAB, Carrer de Can Magrans, E-08193, Barcelona, Spain}

\author{A.~Riggio}
\affiliation{Dipartimento di Fisica, Universit\`a di Cagliari, SP Monserrato-Sestu, Km 0.7, I-09042 Monserrato, Italy}

\author{S.~Serrano}
\affiliation{Institute of Space Sciences (ICE, CSIC​)​, Campus UAB, Carrer de Can Magrans, E-08193, Barcelona, Spain}

\author{A.~Veledina} 
\affiliation{Department of Physics and Astronomy, FI-20014 University of Turku, Finland}
\affiliation{Nordita, KTH Royal Institute of Technology and Stockholm University, Roslagstullsbacken 23, SE-10691 Stockholm, Sweden}
\affiliation{Space Research Institute of the Russian Academy of Sciences, Profsoyuznaya Str. 84/32, Moscow, 117997, Russia}

\author{L.~Zampieri}
\affiliation{INAF--Osservatorio Astronomico di Padova, Vicolo dell'Osservatorio 5, I-35122 Padova, Italy}



\begin{abstract}

  {\1023} is the first millisecond pulsar discovered to pulsate in the
  visible band; such a detection took place when the pulsar was
  surrounded by an accretion disk and also showed X-ray pulsations.
  We report on the first high time resolution observational campaign
  of this transitional pulsar in the disk state, using
    simultaneous observations in the optical ({\it TNG}, {\it NOT},
    {\it TJO}), X-ray ({\it XMM-Newton}, {\it NuSTAR}, {\it NICER}),
    infrared ({\it GTC}) and UV ({\it Swift}) bands. Optical and
  X-ray pulsations were detected simultaneously in the X-ray {\it
    high} intensity mode in which the source spends $\sim 70\%$ of the
  time, and both disappeared in the {\it low} mode, indicating a
  common underlying physical mechanism. In addition, optical and X-ray
  pulses were emitted within a few km, had similar pulse shape and
  distribution of the pulsed flux density compatible with a power-law
  relation $F_{\nu}\propto \nu^{-0.7}$ connecting the optical and the
  0.3--45~keV X-ray band.  Optical pulses were detected also during
  {\it flares} with a pulsed flux reduced by one third with respect to
  the {\it high} mode; the lack of a simultaneous detection of X-ray
  pulses is compatible with the lower photon statistics. We show that
  magnetically channeled accretion of plasma onto the surface of
    the neutron star cannot account for the optical pulsed luminosity
  ($\sim 10^{31}$~erg~s$^{-1}$).  On the other hand,
  magnetospheric rotation-powered pulsar emission  would require
    an extremely efficient conversion of spin-down power into pulsed
  optical and X-ray emission.  We  then propose that optical and
  X-ray pulses are  instead produced by synchrotron emission from
  the intrabinary shock that forms where a striped pulsar wind meets
  the accretion disk, within a few light cylinder radii away, $\sim
  100$~km, from the pulsar.

\end{abstract}


\section{Introduction} \label{sec:intro}

Transitional millisecond pulsars are fast spinning, weakly magnetic
($B_*\approx 10^8$~G) neutron stars (NSs) with a low mass ($\simlt
M_{\odot}$) companion star, that undergo transitions between distinct
emission regimes over a timescale of less than a couple of weeks. During
bright X-ray outbursts ($L_X\simgt 10^{36}$~erg~s$^{-1}$) they behave
like accreting millisecond pulsars (\citealt{wijnands1998}; see
\citealt{patruno2012,campana2018} for reviews), which accrete matter
transferred by the donor through a disk and emit X-ray pulsations due
to the channeling of the plasma in-flow onto the magnetic poles. When
accretion stops ($L_X\simlt 10^{32}$~erg~s$^{-1}$) they behave as
redback pulsars \citep{damico2001,roberts2018}; the rotation of the NS
magnetic field powers both radio and high energy (X-rays, gamma-rays)
pulsed emission, and a relativistic wind that shocks off the matter
transferred by the companion close to the inner Lagrangian point of
the binary, and ejects it from the system \citep[see,
  e.g.,][]{burderi2001}. {\igr}/PSR M28-I performed a clear transition
between these two regimes in 2013 \citep{papitto2013,ferrigno2014}.

State transitions from two more millisecond pulsars have been
observed, so far, {\1023}
\citep{archibald2009,stappers2014,patruno2014} and {\xss}
\citep{demartino2010,demartino2013,bassa2014}. However, the accretion
disk state of these sources was peculiar, it lasted almost a decade
and had an X-ray luminosity much lower ($L_X\simeq 5
\times10^{33}$~erg~s$^{-1}$) than that usually shown by low mass X-ray
binaries \citep{linares2014}. The X-ray light curve repeatedly showed
transitions on a time scale of $\sim 10$~s between two intensity modes
characterized by a definite value of the luminosity; these were dubbed
{\it high} ($L_X\simeq 7 \times10^{33}$~erg~s$^{-1}$) and {\it low}
($L_X\simeq 10^{33}$~erg~s$^{-1}$) modes
\citep{bogdanov2015,campana2016}. X-ray flares reaching up to a few
times $10^{34}$~erg~s$^{-1}$ were also observed. Coherent X-ray
pulsations were detected only during the {\it high} mode
\citep{archibald2015,papitto2015} and interpreted in terms of
magnetically channeled accretion onto the magnetic poles, even though
the low X-ray luminosity at which they were observed should not allow
matter to overcome the centrifugal barrier due to pulsar rotation
\citep[but see also][]{bozzo2018}. Moreover, the spin-down rate of {\1023}
during the disk state \citep[][J16 in the following]{jaodand2016} was
close to the value taken during the rotation-powered radio pulsar
state and its modulus is lower by at least one order of magnitude than
that expected if accretion or propeller ejection of matter takes
place.  A relatively bright continuous radio emission with a flat
spectral shape was detected and interpreted as compact self-absorbed
synchrotron jet \citep{deller2015}, with radio flares occurring during
the X-ray {\it low} mode indicating ejection of optically thin
plasmoid \citep{bogdanov2018}.  Flares and flickering reminiscent of
the {\it high}/{\it low} mode transitions were also seen in the
optical \citep{shahbaz2015,papitto2018,kennedy2018} and in the
infrared \citep{hakala2018,shahbaz2018} bands; the optical light was
polarized at a $\sim 1\%$ degree \citep{baglio2016}, to a different
extent during the various modes \citep{hakala2018}.  The appearance of
a disk in transitional millisecond pulsars in such a sub-luminous
state was also accompanied by a factor of a few increase in the GeV
gamma-ray luminosity \citep{torres2017}, while only upper limit were
placed on the TeV flux \citep{veritas}.

This complex phenomenology led to a flurry of different
interpretations either based on the emission of a rotation powered
pulsar enshrouded by disk matter \citep{takata2014, CZ2014, li2014}, a pulsar
that propels away disk matter \citep{papitto2014,papitto2015b} or a
pulsar accreting mass at a very low rate from disk trapped near
corotation \citep{dangelo2012}. The possibility that switching between
the {\it high} and the {\it low} X-ray modes marked changes between a
radio pulsar and a propeller regime was  also proposed
\citep{linares2014,campana2016,cotizelati2018}.

Recently, \citet{ambrosino2017} discovered optical pulsations at the
spin period of {\1023}, produced by a region a few tens of km away
from the NS. This finding was interpreted by the authors as a strong
indication that a rotation-powered pulsar was active in the system,
even when the accretion disk was present. In fact, the pulsed
luminosity observed in the visible band was too large to be produced
by reprocessing of accretion-powered X-ray emission or cyclotron
emission by electrons in the accretion columns above the pulsar polar
caps. Here we report on the first high time resolution
multi-wavelength observational campaign of {\1023} in the disk state
aimed at exploring the relation between optical and X-ray pulses, and
their properties in the different intensity modes. The campaign was
performed on May 23-24, 2017 and involved simultaneous high time
resolution observations by: the fast optical photometer {\it SiFAP}
mounted at the {\it INAF Telescopio Nazionale Galileo} ({\tng}), X-ray
instruments on-board {\xmm} and {\nustar}, and the {\it Canarias
  InfraRed Camera Experiment} (CIRCE) at the {\it Gran Telescopio
  Canarias} (GTC). Optical, UV and X-ray spectra and images were also
obtained thanks to {\not} (NOT), {\tjo} (TJO) and {\swfirst}
observations. Further high time resolution optical observations were
performed by {\tngsifap} on December 20, 2017, and were analyzed
with observations performed by the X-ray {\it NICER} mission a few
weeks earlier.

\section{Observations} \label{sec:obs}

Tab.~\ref{tab:log} lists the observations analyzed and discussed in
this paper, we give details on the analysis of the different data sets
in the following.

\begin{deluxetable*}{lrrcc}[b!]
  \tablecaption{Log of the observations of {\1023}}
  \tablecolumns{5}
\tablehead{
\colhead{Telescope/Instrument} &
\colhead{MJD start time\tablenotemark{a}} &
\colhead{Exposure (s)} &
\colhead{Band} &
\colhead{Mode/Magnitude} \\
}
\startdata
2017 May 23 & & & & \\
{\nustar}/FPMA-FPMB & 57896.035995 & 82514.0 & 3--79~keV&  \\
{\not}/ALFOSC & 57896.896694 & 3600.0 & 440--695~nm & grism\#19 \\
    {\xmm}/MOS1 (XMM1)   & 57896.905272 & 24651.7 & 0.3--10~keV & Small window \\
    {\xmm}/MOS2 (XMM1)   & 57896.905765 & 24613.9 & 0.3--10~keV & Small window \\
       {\tjo}/MEIA2 & 57896.907924 & 300.0 & Johnson-V & $16.85\pm0.01$ \\
       {\sw}/UVOT & 57896.915625 & 1710.7 & UVW1 & Imaging+Event \\
        {\sw}/XRT & 57896.915579 & 1721.6 & 0.3--10~keV & Photon counting \\
        {\xmm}/EPIC-pn (XMM1) & 57896.929398 & 24914.0 & 0.3--10~keV & Timing \\
        {\it GTC}/CIRCE & 57896.930555 & 4800.0 & {\it Ks} & Fast imaging \\
    {\tngsifap} (TNG1)& 57896.970058 & 3297.7  & white filter& Fast timing \\
    \hline
    2017 May 24 & & & & \\
    {\xmm}/EPIC-pn (XMM2) & 57897.739274 & 23413.0 & 0.3--10~keV & Timing \\
    {\xmm}/MOS1 (XMM2)   & 57897.715133 & 23150.6 & 0.3--10~keV & Small window \\
    {\xmm}/MOS2 (XMM2)   & 57897.715642 & 23112.8 & 0.3--10~keV & Small window \\
    {\tngsifap} (TNG2)& 57897.890802 & 8397.2  & white filter& Fast timing \\
        {\tjo}/MEIA2 & 57897.896660  & 600.0 & U-Johnson & $18.1\pm0.1$ \\
        {\tjo}/MEIA2 & 57897.903954  & 200.0 & B-Johnson & $17.10\pm0.03$ \\
        {\tjo}/MEIA2 & 57897.906622  & 200.0 & V-Johnson & $16.64\pm0.02$ \\
            {\tjo}/MEIA2 & 57897.909287 & 200.0 & R-Cousins & $16.31\pm0.02$ \\
            {\tjo}/MEIA2 & 57897.911955 & 200.0 & I-Cousins & $15.91\pm0.02$ \\
            \hline
            2017 December 2--20 & & & & \\
            {\nicer} (1034060101) & 58089.019028 &  3070.0 & 0.2--12~keV &  \\
            {\nicer} (1034060102) & 58090.044676 &  1963.0 & 0.2--12~keV&  \\
            {\nicer} (1034060103) & 58093.393750 &  1144.0 & 0.2--12~keV&  \\      
      {\tngsifap} (TNG3-B)& 58107.126433 & 3298.3 & Johnson-B & Fast timing \\
         {\tngsifap} (TNG3-V)& 58107.179563 & 3298.3 & Johnson-V & Fast timing \\
         {\tngsifap} (TNG3-R)& 58107.235471 & 3298.3 & Johnson-R & Fast timing \\
    \enddata
\tablenotetext{a}{Barycentric Dynamical Time at exposure start.}
\label{tab:log}
\end{deluxetable*}

\subsection{X-ray observations}

\subsubsection{{\xmm}/EPIC}
\label{sec:xmm}

We analyzed {\xmm} Target of Opportunity (ToO) observations of {\1023}
performed on 2017, May 23 (Id. 0794580801, XMM1 in the following) and
May 24 (Id. 0794580901, XMM2 in the following) in the Discretionary
Time of the Project Scientist (PIs: Papitto, Stella). We used the SAS
(\texttt{Science Analysis Software}) v.16.1.0 to reduce the data.  We
transformed the photon arrival times observed by {\xmm} as if they
were observed at the line of nodes of the Solar System barycenter
using the source position derived by \citet{deller2012},
RA=10:23:47.687198(2), DEC=00:38:40.84551(4), the JPL ephemerides
DE405 and the \texttt{barycen} tool. We used the same parameters to
correct arrival times observed by other instruments considered in this
paper. We discarded the first 3.5~ks of XMM1 data from the analysis
because high flaring particle background -- apparent from the
10-12~keV light curve -- contaminated data and prevented
identification of X-ray modes. In both observations the EPIC-pn was
operated with a time resolution of 29.5~$\mu$s (timing mode) and a
thin optical blocking filter.  We defined source and background
regions with coordinates RAWX=27--47 and RAWY=3--5, respectively, and
retained good events characterized by a single or a double
pattern. EPIC MOS1 and MOS2 cameras observed the target in small
window mode with a time resolution of 0.3~s, and a thin optical
blocking filter. We extracted source photons falling within a circular
region centered on the source position with a 35\arcsec radius, and
background photons from a 100\arcsec-wide, source-free circular region of
one of the outer CCDs; we retained good events with patterns as
complex as quadruples. We created background-subtracted light curves
with the task \texttt{epiclccorr}. Redistribution matrices and
ancillary response files were computed using \texttt{rmfgen} and
\texttt{arfgen}, and the spectra re-binned to have at least 25 counts
per bin and no more than three bins per resolution element in the
0.3--10~keV energy band.

\subsection{\nustar}

We analyzed the {\nustar} ToO observation of {\1023} performed on
2017, May 23 (Id. 80201028002, starting at 00:26:09 UTC and lasting
160~ks, for a total exposure time of 82.5~ks; PI: Papitto). We reduced
the observation by performing standard screening and filtering of the
events with the {\nustar} data analysis package (NUSTARDAS) version
v.1.8.0 with CALDB 20170126. We selected source and background events
from circular regions of 55\arcsec radius centered at the source
location and in a source-free region away from the source, respectively.

\subsection{\nicer}

We present the analysis of {\nicer} observations of {\1023} performed
on 2017, December 2 (Id. 1034060101), December 3 (Id. 1034060102) and
December 6 (Id. 1034060103). The events across the 0.2-12~keV band
\citep{Gendreau2012} were processed and screened using the HEASOFT
version 6.24 and NICERDAS version 4.0.

\subsection{{\sw}/XRT}

We consider the {\sw} observations of {\1023} that started on 2017,
May 23 at 21:53 (UT, Id. 00033012149) with an exposure of 1.7~ks. We
reduced data obtained with the {\it X-ray Telescope} (XRT) in Photon
Counting mode using the HEASoft tool {\texttt{xrtpipeline}}, extracted
light curves and spectra with {\texttt{xselect}} from a circle with a
47\arcsec radius centered on the source position, and produced
ancillary response files using {\texttt{xrtmkarf}}. The XRT observed a
variable count rate between 0.05 and 0.4~s$^{-1}$. The 0.3--10~keV
spectrum could be described by an absorbed power law with absorption
column fixed $N_H=5\times10^{20}$~cm$^{-2}$ and photon index
$\Gamma=1.6\pm0.2$, giving an unabsorbed 0.3--10~keV flux of
$(1.0\pm0.1)\times10^{-11}$~erg~cm$^{-2}$~s$^{-1}$, typical for the
disk state of {\1023}.

\subsection{Optical/UV observations}

\subsection{\tng}
\label{sec:tngobs}

We observed {\1023} with the {\sifap} fast optical photometer
\citep{meddi2012,ambrosino2016,ambrosino2017} mounted at the 3.6~m TNG
starting on 2017, May 23 at 23:17 (TNG1, overlapping for 3.0~ks with
XMM1), on 2017, May 24 at 21:21 (TNG2, overlapping for 8.0~ks with
XMM2), and on 2017, December 20 at 03:02 (TNG3). The three
observations were performed during the Director discretionary time.
We observed {\1023} using the channel of SiFAP that ensured the
maximum possible time resolution (25~ns), and a $V=15.6$~mag reference
star UCAC4 454-048424 \citep{zacharias2013} with the second channel
which operated with a time resolution of 1~ms during TNG1, and 5~ms
during TNG2 and TNG3.   The size of the on-source channel of SiFAP
is $0.13\times0.13\,cm^2$ which at the $F/11$ focus of the TNG
corresponds to nearly $7\times7\,$arcsec$^2$. In TNG1 and TNG2 we
  used a white filter covering the 320--900~nm band and maximum
  between 400 and 600~nm (i.e. roughly corresponding to the B and V
  Johnson filters, see Supplementary Fig.1 in
  \citealt{ambrosino2017}). During TNG3 we observed the source with
  Johnson {\it B} ($\lambda_{eff}=445$~nm,
  $\Delta\lambda_{FWHM}=94$~nm), {\it V} ($\lambda_{eff}=551$~nm,
  $\Delta\lambda_{FWHM}=88$~nm) and {\it R} ($\lambda_{eff}=658$~nm,
  $\Delta\lambda_{FWHM}=138$~nm) filters, for 3.3~ks each; filters
  could not be used on the reference star due to problems with the
  instrumental set-up . We estimated the background by tilting the
  pointing direction of the telescope by a few tens of arcseconds for
  $\sim 100$~s; this was done four times during TNG2 (roughly every
  half an hour) and once during TNG1 (at the end of the exposure) and
  during each of the three filtered observations of TNG3. The
  background count rate increased during TNG2 as the elevation of the
  source over the horizon decreased and the contamination by diffuse
  light was correspondingly larger; we then evaluated the background
  contribution by fitting the count rate observed during the four
  intervals with a quadratic polynomial. During TNG1 the background
  was estimated only at the end of the exposure, and considering that
  its contribution increases as the source declines over the horizon,
  it is almost certainly larger than the actual value that affected
  the observation of the source.

The SiFAP clock showed drifts with respect to the actual time measured
by two  Global Positioning System (GPS) pulse-per-second (PPS)
  signals that were used to mark the start and the stop times of each
  observation. For this reason the total elapsed time by the clock
  exceeded the GPS time by $\Delta t_{SiFAP}-\Delta t_{GPS}=0.84$ and
  $4.23$~ms during TNG1 and TNG2, each lasting $\Delta t_{GPS}=3300.0$
  and 8400.0~s, respectively. During TNG3, the SiFAP clock lagged the
  GPS signal by $3.89$, $3.13$ and $3.08$~ms during the exposures with
  B, V and R filters, respectively, each lasting $\Delta
  t_{GPS}=3300.0$~s. As no further information on the dependence of
  the drift on the various parameters that affect the photometer
  operations (e.g. temperature, count rate) is available, the best
  possible guess is that the drift rate was constant. Following
  \citet{ambrosino2017}, we corrected the arrival times recorded by
  SiFAP using the relation $t_{arr}=t_{SiFAP}\times(\Delta
  t_{GPS}/\Delta t_{SiFAP})$. Subsequently we used the software
  {\texttt{tempo2}} \citep{hobbs2006} to correct the photon arrival
  times to the Solar System barycenter, using the position of {\1023}
  reported in Sec.~\ref{sec:xmm} and the geocentric location of the
  TNG (X=5327447.481, Y=-1719594.927, Z=3051174.666), along with the
  JPL ephemerides DE405. During the 3.3~ks exposure of TNG1 we measured
  an average count rate of $\sim 2.59\times10^4$, $3.70\times10^4$ and
  $1.75\times10^4$~s$^{-1}$ from the source, the reference star and
  the sky background, respectively. Values of $2.95\times10^4$,
  $4.19\times10^4$ and $1.11\times10^4$~s$^{-1}$ were measured for the
  same quantities during TNG2.

\subsection{{\xmm}/OM}

The Optical Monitor (OM) on-board {\xmm} observed the source during
XMM1 and XMM2 with a B filter ($\lambda_{eff}=450$~nm) and using the
fast mode, which has a time resolution of 0.5~s. We extracted source
photons from a 6-pixel wide circle (corresponding to 2.9~arcsec) and
background from an annulus with inner and outer radius of 7.2 and 9.9
pixel, respectively. Observed count rates, $R_{OM}$, were converted to
flux and magnitude units using the relations\footnote{see
  \url{https://www.cosmos.esa.int/web/xmm-newton/sas-watchout-uvflux}. }
$f_{\lambda}= 1.29\times10^{-16}\,
R_{OM}$~erg~cm$^{-2}$~s$^{-1}$~{\AA}$^{-1}$ and $mag =
19.266-2.5\log_{10}R_{OM}$.
%

\subsection{{\sw}/UVOT}

The Ultraviolet Optical Telescope (UVOT) on-board {\sw} observed
{\1023} with the UVW1 filter ($\lambda_c=260$~nm) for 674.1~s,
starting on May 23, 2017 at 21:58. We performed aperture photometry by
using a circle with radius of 5~arcsec around the source position, and
extracting the background from a nearby source-free region. We
measured an average flux of $16.05\pm0.03$(stat)$\pm0.03$(sys)~mag
(Vega system) from the source with the tool {\texttt{uvotsource}},
corresponding to a flux density of
$(1.5\pm0.1)\times10^{-15}$~erg~cm$^{-2}$~s$^{-1}$, or
$(3.38\pm0.21)$~mJy.


\subsection{\not}
\label{sec:not}

We performed a ToO target of opportunity observation of {\1023} (PI:
Papitto) with the 2.5~m NOT starting on 2017, May 23 (21:28 UTC),
taking four 900~s long spectra with the Andalucia Faint Object
Spectrograph and Camera (ALFOSC), equipped with the 0.5~arcsec slit
(appropriate to the actual seeing of 0.7~arcsec), and  grism \#19
(440-695~nm). Subsequently five photometric images were taken with the
SDSS {\it u'}, {\it g'}, {\it r'}, {\it i'} and {\it z'} filters with
an exposure of 60~s (except for a 120~s exposure for the the {\it u'}
image).

We reduced data using standard \texttt{IRAF} \citep{tody1986} tasks
such as bias and flat-field correction, cosmic ray cleaning,
wavelength calibration, extraction of spectra from science frames
using the optimized method by Horne (1986) and flux calibration. The
wavelength calibrations made use of He and Ne arc lamps as a
reference, while the flux calibration was made using the simultaneous
observation performed by TJO in the Johnson V-band
($V=16.85\pm0.02$~mag, corresponding to
$F_{\lambda}=(6.9\pm0.1)\times10^{-16}$~erg~cm$^-2$~s$^{-1}$~\AA$^{-1}$,
see \ref{sec:tjo}), and the La Palma standard extinction curve. The
spectrum was first extracted for each differential image, then, the
four extracted spectra were combined together.

We used \texttt{Starlink GAIA} v.~4.4.8 to perform aperture photometry
on the five images taken by ALFOSC, using the nearby stars
UCAC4~454-048421 and UCAC4~454-048418 \citep{zacharias2013} to
calibrate the magnitude scale, and an aperture between 3 and 5 arcsec
depending on the filter, and obtained the following magnitudes,
$u'=17.43(1)$, $g'=16.836(4)$, $r'=16.625(4)$, $i'=16.488(5)$,
$z'=16.370(8)$.

\subsection{\tjo}
\label{sec:tjo}
TJO is a robotic 80~cm telescope located at the Observatori
Astron\`omic del Montsec (Catalunya, Spain). We observed the field
around {\1023} on 2017, May 23 and 24 (see Table~\ref{tab:log} for
details), in the context of an observing program to monitor
transitional millisecond pulsars (Id. p153, PI: Papitto). We used the
photometric imaging camera MEIA2, equipped with Johnson U, B and V
filters and Cousins R and I filters. We created average dark and bias
frames and flat-fielded images using the ESO \texttt{eclipse} package
v.~5.0-0\footnote{Available at
  http://www.eso.org/sci/software/eclipse/} and performed aperture
photometry with the \texttt{Starlink GAIA} v.~4.4.8 using an aperture
of 12~pixel (equivalent to 4.3~arcsec). We used the same nearby
reference stars used to calibrate {\it NOT} images (see
Sec.~\ref{sec:not}), and obtained magnitudes reported in the rightmost
column of Table~\ref{tab:log}.

\subsection{\it Gran Telescopio Canarias}

Fast near-Infrared imaging of {\1023} was carried out with a ToO
observations (PI: Rea) on the night of 2017 May 23 using {\it Canarias
  InfraRed Camera Experiment} (CIRCE; Eikenberry et al. 2017) on the
10.4~m GTC.  We configured the detector in fast imaging mode using a
window size of 2048x1366 with the plate scale, 0.1 arcsec/pixel.  We
obtained 475 images with 4.9 seconds exposure time using Ks filter
from 57896.93 to 57896.98 MJD.  Due to fast variations in the infrared
sky, we dithered the telescope in every five images with a five-point
dither pattern.  Appropriate flat and dark frames were obtained during
the twilight and at the end of the night respectively.

Data reduction of the CIRCE data was carried out using SuperFATBOY
data reduction pipeline.  We applied the standard procedures in the
following order; linearity correction, dark subtraction and dividing
by master flat.  Infrared sky background was obtained separately for
each dither pattern consisting of 25 images.  In the end, we
interpolated over bad pixels and cosmic ray events, and binned the
images by two pixels to improve signal to noise ratio.

Extracting the photometry, we roughly aligned all the images, then, we
further adjusted the location of each source ({\1023} and 3 reference
stars) with a Gaussian fit.  We extracted the source counts from an
aperture of 8 pixel radius and determined the sky background from an
annulus with 1.5 to 2 times the aperture size.

\section{Data Analysis} \label{sec:analysis}

\subsection{Light Curves}

\subsubsection{X-ray light curve}

\label{sec:xraylc}

\begin{figure*}[t!]
\includegraphics{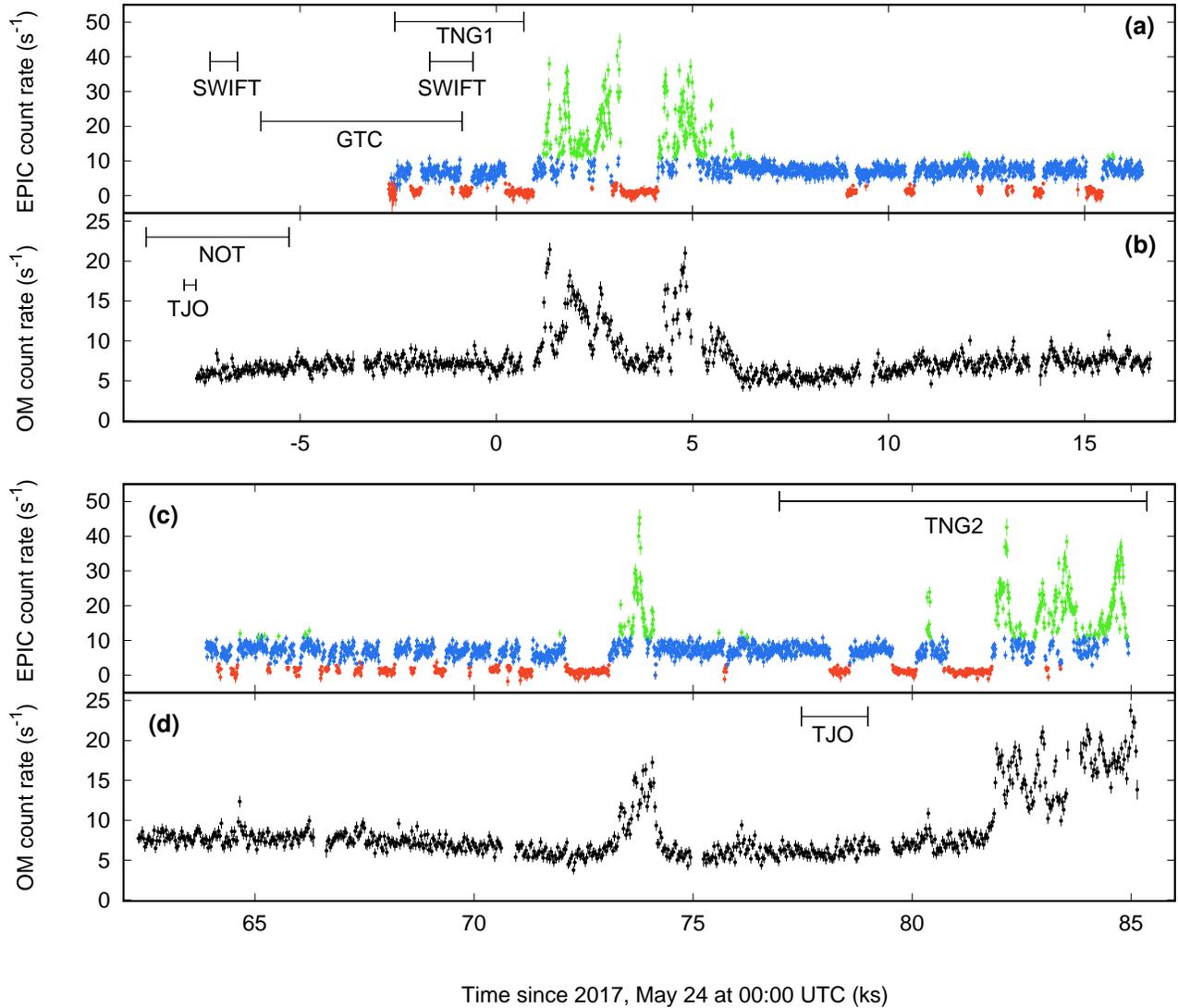}
 \caption{0.3--10~keV EPIC light curve observed during XMM1 (panel a)
   XMM2 (panel c), binned every 10~s. Blue, red and green points
   indicate {\it high}, {\it low} and {\it flaring} mode intervals
   (see text for their definition).  Panel (b) and (d) show the light
   curve observed by the {\it Optical Monitor} on board {\xmm} and
   binned every 30~s. Horizontal bars indicate the intervals of
   simultaneous observations performed by other instruments.}
 \label{fig:lc1}
 \end{figure*}

We built an EPIC 0.3--10~keV light curve with a time resolution of
10~s, summing the background-subtracted light curves of the three EPIC
cameras, pn, MOS1 and MOS2.  We adopted the definition of X-ray modes
of \citet{bogdanov2015} and considered {\it low} mode intervals with a
count rate lower than 3.1~s$^{-1}$, {\it flaring} intervals with a
count rate larger than 11~s$^{-1}$, and {\it high} mode the intervals
with a count rate in between these thresholds. We also adopted the
bi-stable comparator defined by \citet{bogdanov2015} and defined an
intermediate {\it gray} area ranging from 2.1 to 4.1~s$^{-1}$; we did
{\it not} flag as a transition between {\it high} mode and {\it low}
mode (or vice versa) when the count rate varied from the {\it high}
mode region to the {\it gray} area and then returned back to the {\it
  high} mode region, but considered the whole interval as {\it high}
mode. Fig.~\ref{fig:lc1} shows the light curves observed during XMM1
(panel a) and XMM2 (panel c), respectively. {\it Low}, {\it high} and
{\it flaring} mode intervals are plotted with red, blue and green
points. Panels (b) and (d) of Fig.~\ref{fig:lc1} show the OM optical
light curve during XMM1 and XMM2, respectively.

\begin{figure*}[t!]
 \includegraphics{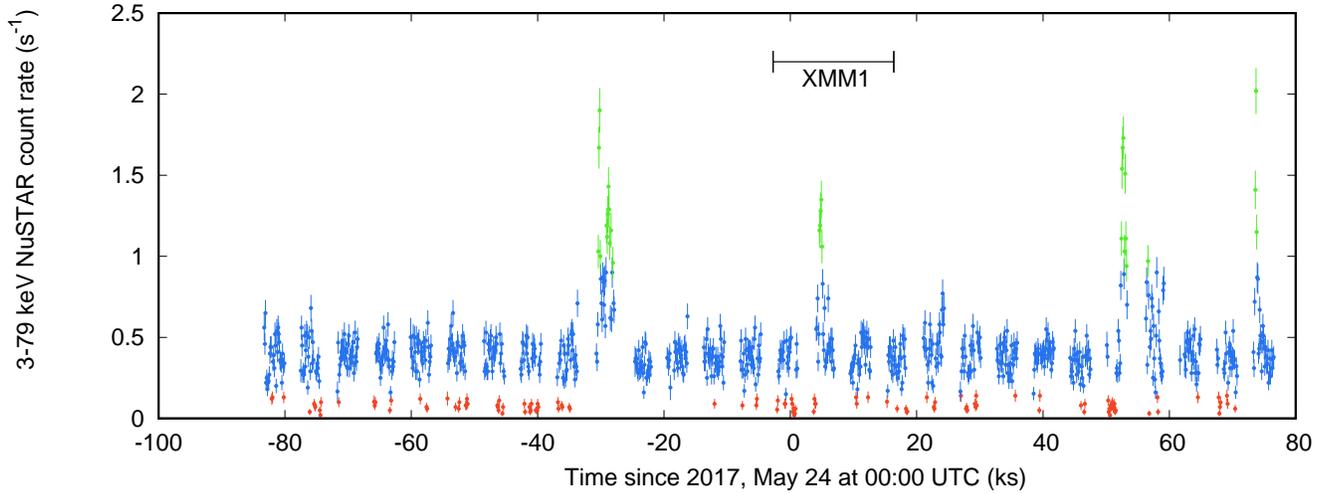}
\caption{3--79 keV light curve observed by {\nustar} binned at 100~s
  intervals. Red, blue and green points mark the {\it low}, {\it high}
  and {\it flaring} mode, respectively (see text for their
  definition). The horizontal bar indicates the interval covered by XMM1 observation.}
 \label{fig:nustar_lc}
 \end{figure*}

Figure~\ref{fig:nustar_lc} shows the sum of the 3--79~keV light curves
observed by two {\nustar} modules, binned at a time resolution of
100~s. We identified the transition using a light curve sampled in
10~s-long time bins, and setting the threshold between the {\it low}
(red points) and the {\it high} (blue points) mode at a count rate of
0.15~s$^{-1}$, whereas the {\it flaring} (green points) took place at
a rate higher than 0.9~s$^{-1}$.  These threshold are similar to those
determined by \citet{tendulkar2014} and \citet{cotizelati2018}; we set
them by requiring that the the modes determined from the {\nustar}
light curve would be the same as those observed by {\xmm} during the
11.1~ks-long overlap occurring with XMM1. Due to the different photons
statistics an exact correspondence could not be found; the threshold
were conservatively set in order to avoid that the {\nustar} intervals
deemed as {\it flaring} or {\it low} would be contaminated by {\it
  high} mode emission.

\subsubsection{Correlated optical/X-ray variability}

 \begin{figure*}[t!]
  \includegraphics{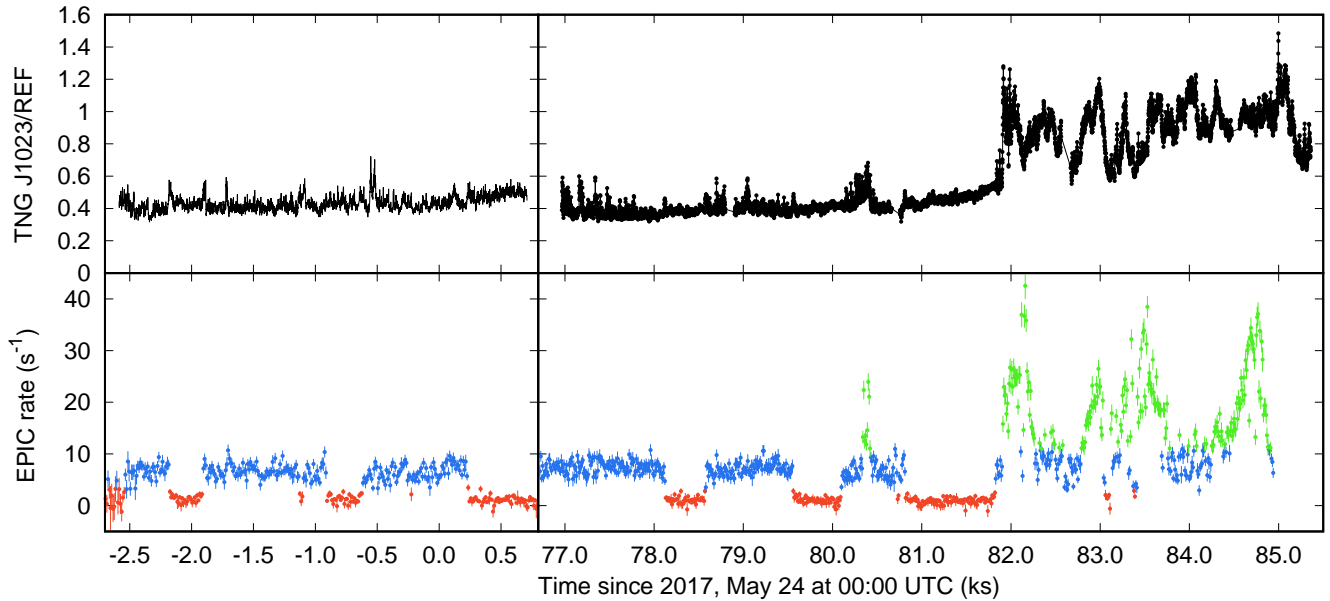}
\caption{Differential {\it TNG}/SiFAP optical light curve of {\1023}
  (top panel) and simultaneous {\xmm}/EPIC coverage (bottom
  panel). Blue, red and green points indicate {\it high}, {\it low}
  and {\it flaring} mode intervals (see text for their definition). }
 \label{fig:lc2}
 \end{figure*}
    
The top panel of Figure~\ref{fig:lc2} shows the differential
{\tngsifap} optical light curve defined as the ratio between the
background subtracted count rate observed from {\1023} and the
reference star employed at the {\it TNG}, together with the
simultaneous {\xmmepic} X-ray light curves in the respective bottom
panels. Both light curves were binned every 10~s. Correlation between
flares is evident, while there is no clear optical analogue of the
{\it high}/{\it low} mode transitions observed in the X-ray band.

\begin{figure}[t!]
\includegraphics[width=8.6cm]{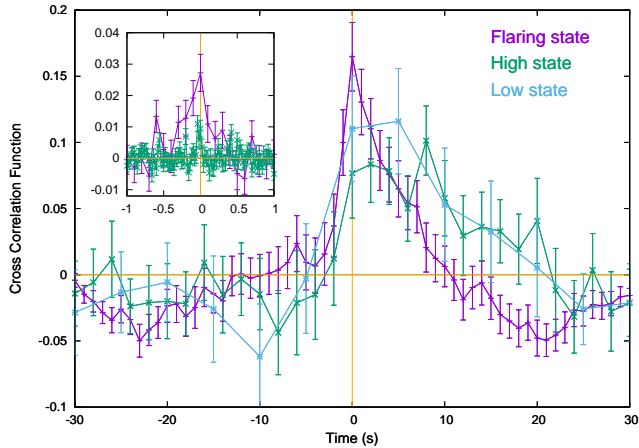}
   \caption{CCFs between the optical ({\tngsifap}) and X-ray
     ({\xmmepic}) light curve observed during the {\it flaring}
     (magenta points), {\it high} (green points) and {\it low} (blue
     points) modes. A time resolution ranging between 1 and 2.5~s was
     used in the different modes, and 64~s-long intervals were
     averaged. The inset shows the CCF evaluated on a shorter time
     scale (ranging between 25 and 100 ms).}
 \label{fig:ccf}
 \end{figure}

In order to explore the degree of correlation between the X-ray and
the optical variability we considered the light curves observed by
{\xmmepic} and {\tngsifap} binned with a shorter timescale. The three
intensity modes defined in Sec.~\ref{sec:xraylc} introduce a strong
non-stationariety in the light curves. The cross-correlation function
(CCF) requires stationarity, therefore we calculated a CCF for each
mode. The intervals were selected based on the X-ray behavior, and
were chosen carefully so as to exclude transitions. The CCFs were
Scalculated using the HEASoft tool \texttt{crosscor} with a time
resolution ranging from 1 to 2.5~s over 64~s intervals of the {\it
  flaring}, {\it high} and {\it low} modes; they are plotted in
Fig.~\ref{fig:ccf} with magenta, green and blue symbols,
respectively. The inset of Fig.~\ref{fig:ccf} shows the CCFs evaluated
on a shorter time scale, ranging from 25 and 100 ms.

Our analysis shows that X-ray and optical emissions were clearly
correlated in each of the three modes, with a somewhat similar
behavior and degree of correlation. Namely, on timescales longer than
a second (see Fig.~\ref{fig:ccf}), the optical variability showed a
range of delays with respect to the X-ray variability, with a
“reflection-like” CCF shape \citep{obrien2002} which rises sharply in
2--3~s, peaks around zero, and decays slowly toward positive delays
(i.e. optical lagging X-rays) reaching $\sim$10-20 seconds. On shorter
timescales (see the inset of Fig.~\ref{fig:ccf}), the optical
variability appeared instead to be  correlated with the X-ray
variability in the {\it flaring} mode with no evidence for delays.
The CCF hints to a correlation also in the {\it high} mode, with
no evidence for delays, while in the {\it low} mode the analysis is
hampered by the lower statistics.

We note however that the shape of the CCFs depends strongly on the
parameters chosen to calculate them, i.e. on the time resolution and
on the length of the intervals over which the Fast-Fourier Transforms
(FFT) were calculated. More importantly, when calculating the CCFs in
the time domain (which by definition implies the use of “non strictly
simultaneous” data intervals), we obtain different results. Finally,
notwithstanding the rather low statistics, we have marginal evidence
for the CCFs being variable in time. If confirmed, all this would
imply that the variability is non-stationary even within a (X-ray)
defined mode.

\subsubsection{Correlated infrared/X-ray variability}

\begin{figure}[t!]
\includegraphics[width=8.6cm]{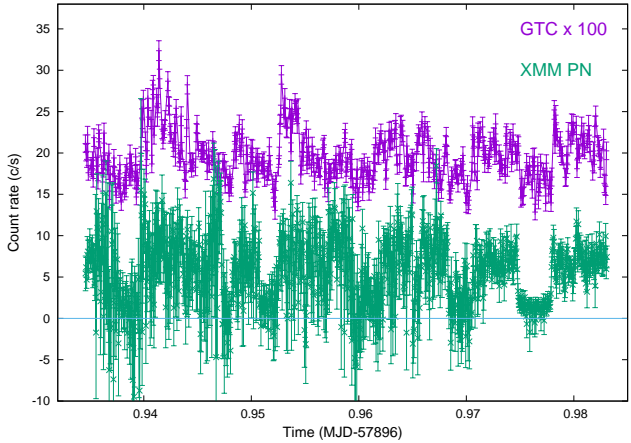}
   \caption{Light curves observed simultaneously by {\it CIRCE/GTC} in
     the {\it Ks} infrared band (magenta points, the observed rates
     were scaled by a factor 100) and by the {\xmmepic} in the
     0.3--10~keV band (green points). A time resolution of 5.88~s was
     used.}
 \label{fig:gtc}
 \end{figure}

The infrared light curve accumulated by {\it CIRCE} at {\it GTC}
overlaps for 4.3~ks with the  EPIC-pn exposure in
XMM1. Fig.~\ref{fig:gtc} shows the light curves observed in the
infrared and X-ray bandwith with magenta and green symbols,
respectively. Note that high particle background flaring affects the
EPIC-pn light curve before MJD~57896.967, i.e. during the first 2.7~ks
of the overlap with the {\it CIRCE} light curve, making it extremely
noisy.  The combination of this high background interval affecting the
X-ray light curve and the short duration and low statistics of the
infrared light curve prevented us to perform a quantitative study of
the correlation between the X-ray and the infrared
variability. However, a visual inspection of the two simultaneous
curves provides clear evidence that they are correlated, with a
slightly lower infrared flux during the X-ray {\it low} mode, when
compared to the {\it high} mode. This is evident especially for the
{\it low} mode occurring close to MJD 57896.976 (see
Fig.~\ref{fig:gtc}).

Additionally, and perhaps more interestingly, there is also possible
evidence for an increase of the infrared flux right after the
transition from the {\it low} to the {\it high} X-ray mode. In other
terms, the start of each X-ray {\it high} mode interval might be
accompanied by a modest infrared flare. However this evidence is based
on the presence of only three of such flares (i.e. those at about
0.94, 0.953 and 0.978~d since MJD~57896.0, the latter also seen in the
TNG optical light curve, see the light curves in the left panel of
Fig.~\ref{fig:lc2} corresponding to -2~ks in the units used there)
while in two additional occasions (0.962 and 0.971) no infrared flare
seems to match the start of an X-ray {\it high} mode interval. Further
and longer simultaneous observations are needed to confirm this
result.

\subsection{Timing Analysis}

\subsubsection{May 2017 campaign. The X-ray pulse}
\label{sec:xmm2017}

\begin{figure}[t!]
\includegraphics[width=8.6cm]{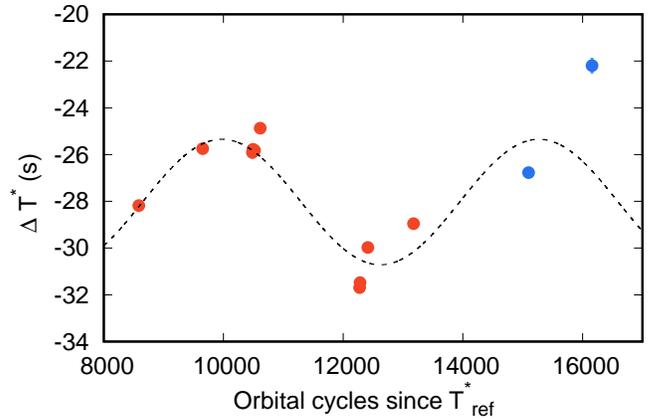}
\caption{Difference between the epoch of passage of the pulsar at the
  ascending node of its orbit measured from X-ray pulsations and those
  expected according to the radio pulsar timing solution measured by
  \citealt{archibald2013} and J16 as a function of the number of
  orbital cycles elapsed since $T^*_{ref}=54905.97140075$.  Red
    symbols refer to epochs measured by J16, blue symbols to measures
    from this work. Uncertainties on each measure range from 0.05 to
    0.3~s.  The dashed line shows a sinusoid with period equal to
  $\approx 5300$ orbital cycles, unable to explain the long-term
  trend.  }
 \label{fig:tstar}
 \end{figure}

In order to perform a search for pulsations in the X-ray datasets we
first corrected for the shifts of the photon arrival time caused by
the orbital motion of the pulsar in the binary system. J16 showed that
the epoch of passage of the pulsar at the ascending node of the orbit
$T^*$ derived from the X-ray pulsations of {\1023} shifts by up to
tens of seconds with respect to any plausible solution based on a
constant orbital period derivative. We then had to determine an
orbital solution valid across the time span by the observations
discussed here. We corrected the two {\xmm}/EPIC-pn time series using
the semi-major axis ($a \sin{i}/c=0.343356$~ lt-s) and orbital period
($P_{orb}=17115.5216592$~s) of the J16 timing solution, and a grid of
values of $T^*$ spaced by 0.125~s around the expected value. We
carried out an epoch folding search on each of the time series by
sampling each period with 16 phase bins, and estimated the best epoch
of passage at the ascending node ($T^*=$57896.82926(1)~MJD) by fitting
with a Gaussian the values of the maximum pulse profile variance found
in each of the periodograms we obtained. Figure~\ref{fig:tstar} shows
the difference $\Delta T^*$ between the epoch predicted by the radio
timing solution of \citet[][see also Table~2 in J16]{archibald2013}
and the values measured from X-ray pulsations as a function of the
number of orbital cycles performed since
$T^*_{ref}=54905.97140075$~MJD. We took from Table~1 of J16 the values
of $\Delta T^*$ measured before $\approx 13200$ cycles had elapsed
since $T^*_{ref}$ (red points in Fig.~\ref{fig:tstar}), and added
the last two measurements based on the analysis presented here (
  blue points in Fig.~\ref{fig:tstar}; see below and
Sec.~\ref{sec:dec2017}). The epoch we measured of passage of the
pulsar at the ascending node in May 2017 anticipated by $\Delta
T_{asc}=$26.8(8)~s the epoch predicted by the radio timing solution,
confirming the $\sim20$--$30$~s shift that occurred after the June
2013 state change already reported in J16, as well as the inability to
describe with a polynomial of low order (or a sinusoidal variation,
see the dashed line in Figure~\ref{fig:tstar}) the evolution of
$T_{asc}$ over the years.
%

\begin{figure}[t!]
\includegraphics{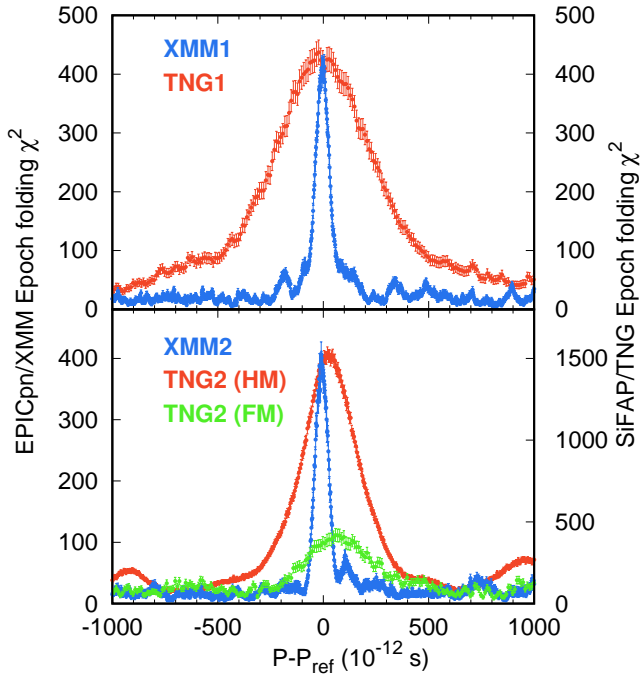}
   \caption{Epoch folding search periodograms of the {\xmm}/EPIC-pn
     (blue points, left scale) and {\tngsifap} (red points, right scale)
     time series in the {\it high} mode (HM), and {\tngsifap} (green points,
     left scale) in the {\it flaring} mode (FM), observed during 2017, May
     23 (top panel) and May 24 (bottom panel). Note that the scale
     used for the periodogram of TNG2 observation is reduced. }
 \label{fig:chicurves}
 \end{figure}

\begin{figure}[t!]
 \includegraphics{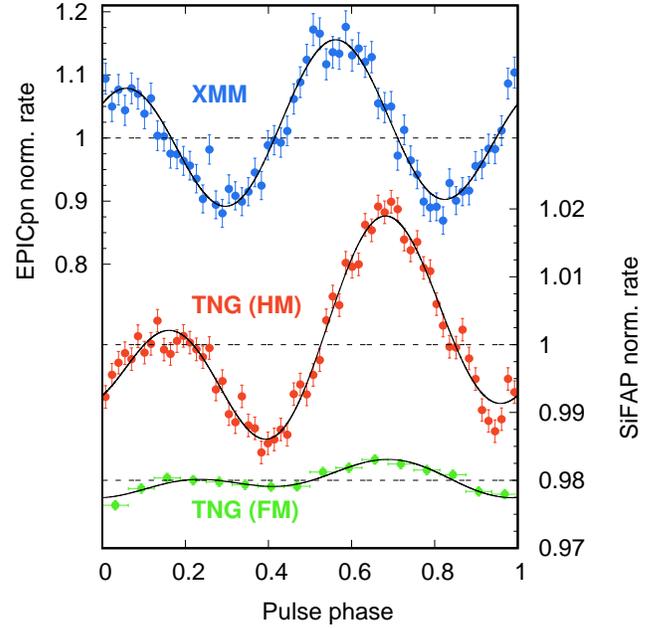}
   \caption{ Pulse profiles obtained by folding the {\it high} mode
     intervals in the {\xmm}/EPIC-pn data (blue points, left y-axis
     scale), the {\it high} mode intervals in the {\tngsifap} data (red
     points, right y-axis scale) and the {\it flaring} mode intervals
     in {\tngsifap} data (green points, right y-axis scale, shifted
     arbitrarily), around $P_{ref}$, with respect to the reference
     epoch MJD $57896.0$.  }
 \label{fig:pulse}
 \end{figure}

In previous studies (\citealt{archibald2015}, J16), X-ray pulsations
were observed only during the {\it high} mode with an RMS amplitude of
$\simeq 8\%$ and no detection was obtained during the {\it low} and
{\it flaring} mode, with an upper limit of 2.4 and 1.4\%,
respectively. We performed a pulsation search on the mode-resolved
EPIC-pn time series and obtained compatible results.  Blue points in
Fig.~\ref{fig:chicurves} show the epoch folding search periodograms
obtained by folding with $n=16$ phase bins the {\it high} mode
intervals observed during XMM1 (top panel) and XMM2 (bottom
panel).  We modeled the folded pulse profiles using two
  sinusoidal harmonics:
  \begin{equation}
    F(\phi)=\sqrt{2}(\bar{F}+\bar{B})\sum_{i=1}^2 r_i \sin[2\pi
      i(\phi-\phi_i)], \end{equation}where $\bar{F}$ and $\bar{B}$ are
  the average source and background count rate, respectively, and
  $r_i$ and $\phi_i$ ($i=1,2$) are the RMS amplitude and phase of the
  two harmonics employed to model the pulse profile.  We estimated
the average period during the two observations by modeling the phases
of the first and second harmonics of the pulse profile computed over
1.2~ks-long intervals using the phase residual formula \citep[see,
  e.g., eq.~1 of][]{papitto2011}, in which we let only the period of
the pulsations and the epoch of passage of the pulsar at the ascending
node free to vary. The second harmonic turned out to have a larger
amplitude and provided the most accurate measurements,
$P_{X}=1.6879874456(5)$~ms, compatible with the period expected
according to the J16 solution, $P_{ref}=1.687987446019(3)$~ms, and
$T^*=57896.829262(1)$~MJD. Blue points in Fig.~\ref{fig:pulse} show
the pulse profiles obtained by folding at $P_{ref}$ the {\it high}
mode intervals of the two EPIC-pn
observations. Table~\ref{tab:periods} lists the background-subtracted
RMS amplitudes $r_i$ and phases $\phi_i$ ($i=1,2$) of the pulse
profile, the total RMS amplitude $R=(\sum_i (r_i^2))^{1/2}$, the
0.3--10~keV flux $F_{abs}$ not corrected for interstellar absorption
and evaluated by modeling the observed spectrum with an absorbed
power law fixing the equivalent hydrogen column density to the value
measured by \citet[][$N_H=5\times10^{20}$~cm$^{-2}$]{CZ2014}, the
isotropic unabsorbed luminosity $L$ evaluated for a distance of
1.37~kpc \citep{deller2012} and the pulsed luminosity
$L_{pulsed}=R\times L$.  Upper limits on the pulse amplitude are
computed at the 95\% confidence level. We also measured the {\it high}
mode RMS amplitude in four energy bands, and used these values to
evaluate the pulsed flux. Cyan and blue points in
Fig.~\ref{fig:spectrum} show the total and the pulsed flux in the {\it
  high} mode, respectively, expressed in $\nu F_{\nu}$ units; a dashed
line indicates a $\nu F_{\nu}^{pulse}\sim\nu^{0.3}$ dependence,
similar to that observed for the total flux.

\begin{deluxetable*}{lccccccccc}[t!]
  \tablecaption{Properties of the X-ray and optical pulses}
  \tablecolumns{3}
\tablehead{
\colhead{Instrument} &
\colhead{Band} &
\colhead{$r_1$ ($\%$)} &
\colhead{$r_2$ ($\%$)} &
\colhead{$\phi_1$} &
\colhead{$\phi_2$}&
\colhead{R ($\%$)}&
\colhead{$F_{abs}$}&
\colhead{$L$}&
\colhead{$L_{pulsed}$}\\
}
\startdata
{\it High} mode & & & & & & & \\
{\xmm}/EPIC-pn & 0.3--10~keV   & $2.9(3)$ & $8.0(3)$ & $0.33(2)$ & $0.434(3)$ & $8.5(4)$ & $13.3(1)$ & 30.5(2) & 2.6(1)\\
{\nustar} \tablenotemark{a} & 3--79~keV &  $<1.6$ & $8.4\pm1.3$ & - & - & $8.4\pm1.3$ & $22(1)$ & 49(2) & 4.1(7) \\
{\it {\tngsifap}} \tablenotemark{b} & 320--900~nm & $0.63(2)$ & $0.74(2)$ & $0.473(7)$ & $0.548(3)$ & $ 0.97(2)$ & $4.2(2)$ & 11.5(4) &  0.111(5) \\
\hline
{\it Low} mode & & & & & & &  \\
{\xmm}/EPIC-pn & 0.3--10 keV & - & - & - & - & $<3.0$ & $2.6(1)$ & 6.1(2) & $<0.18$ \\
{\nustar} \tablenotemark{a} & 3--79~keV & - & - & - & - & $<14.0$ & $2.4(4)$ & 5.4(9) & $<0.75$ \\
{\it {\tngsifap}} & 320--900~nm & $<0.03$ & $<0.02$ & - & - & $<0.034$ & $4.4(2)$ & 12.1(4) & $<0.004$ \\
\hline
{\it Flaring} mode  & & & & & & & \\
{\xmm}/EPIC-pn & 0.3--10 keV & - & - & - & - & $<1.3$ & $37.6(5)$ & 87.1(9) & $<1.1$\\
{\nustar} \tablenotemark{a} & 3--79~keV & -  & - & -  & - & $<6.9$ & $61(8)$ & $137\pm18$ & $<9.5$\\
{\it {\tngsifap}} & 320--900~nm & $0.12(2)$ & $0.11(2)$ & $0.37(2)$ & $0.58(1)$ & $0.16(2)$ & $8.5(4)$ & 23(1) & 0.037(5)\\
\enddata
\label{tab:periods}
 \tablecomments{Pulse profiles were obtained folding the respective
   time series in 64 phase bins (reduced to 16 in case of a low
   significance detection or to derive upper limits) around
   $P_{ref}=1.687987446019$~ms and setting $T_{ref}=57896.0$~MJD as
   the reference epoch. $r_i$ and $\phi_i$ are the
   background-subtracted RMS amplitude and phase (in cycles),
   respectively, of the first ($i=1$) and second ($i=2$) harmonic used
   to model the pulse profiles, $R=(r_1^2+r_2^2)^{1/2}$ is the total
   RMS amplitude, $F_{abs}$ is the flux in units of $10^{-12}$~{\flux}
   not corrected for interstellar absorption and observed in the bands
   listed in the second column, $L$ is the isotropic luminosity in
   units of $10^{32}$~erg~s$^{-1}$ corrected for interstellar
   absorption and evaluated for a distance of 1.37~kpc
   \citep{deller2012}, and $L_{pulsed}=R\times L$ is the pulsed
   luminosity evaluated using the same parameters.  }
 \tablenotetext{a}{{\nustar} pulse profiles were folded around
   $P_{\nustar}=1.687987299(1)$~ms, which differs significantly from
   $P_{ref}$ due to {\nustar} clock drifts. Phases are not reported as
   the {\nustar} clock drifts prevent to draw a meaningful comparison
   with those measured with the other instruments.}
 \tablenotetext{b}{ In TNG2, we considered only the {\it high}
     mode intervals before the onset of the long {\it flaring} event
     which started $\sim$82~ks since May 24 at 00:00 (see
     Fig.~\ref{fig:lc2} and text for details).}
\end{deluxetable*}

We searched the 3--80 keV {\nustar} time series for a coherent signal,
computing power density spectra over 3.3~ks intervals. The average of
the 28 power spectra so obtained shows an excess centered at
1184.8430(1)~Hz interpreted as the second harmonic of the signal at
the spin period of {\1023}. The peak in the power density spectrum is
broad with a width compatible with the spurious derivative of
$10^{-10}$~Hz/s introduced by the {\nustar} clock drift
\citep{madsen2015}, and corresponds to an rms pulse amplitude of
$(6.8\pm1.3)$\%. To determine a more precise value of the pulsation
period we performed an epoch-folding search of the whole {\nustar}
data around the corresponding fundamental frequency obtained from the
analysis of the power density spectrum with steps of $3\times
10^{-14}$~s, for a total of 10001 steps.  The pulse profile with the
largest signal-to-noise ratio corresponds to the period
$P_{NuSTAR}=1.687987299(1)$~ms that differs by $1.5\times10^{-10}$~s
from $P_{ref}$.  We also searched for a signal in the time series
restricted to the three modes. The signal was detected in the {\it
  high} mode with an overall rms amplitude of $(8.4\pm1.3)\%$, roughly
constant up to 45~keV. The pulsed flux spectral distribution (see
green points in Fig.~\ref{fig:spectrum}, where the total flux is also
plotted with light green points) observed by {\nustar} in the high
mode is compatible with the power law relation $\nu F_{\nu} \sim
\nu^{0.3}$ indicated by {\xmm} data. On the other hand, upper limits
of $14.0\%$ and $6.9\%$ (95\% confidence level) were set on the pulse
amplitude during the {\it low} and the {\it flaring} mode,
respectively.

\subsubsection{May 2017 campaign. The optical pulse}
\label{sec:optmay2017}

\begin{figure}[t!]
\includegraphics{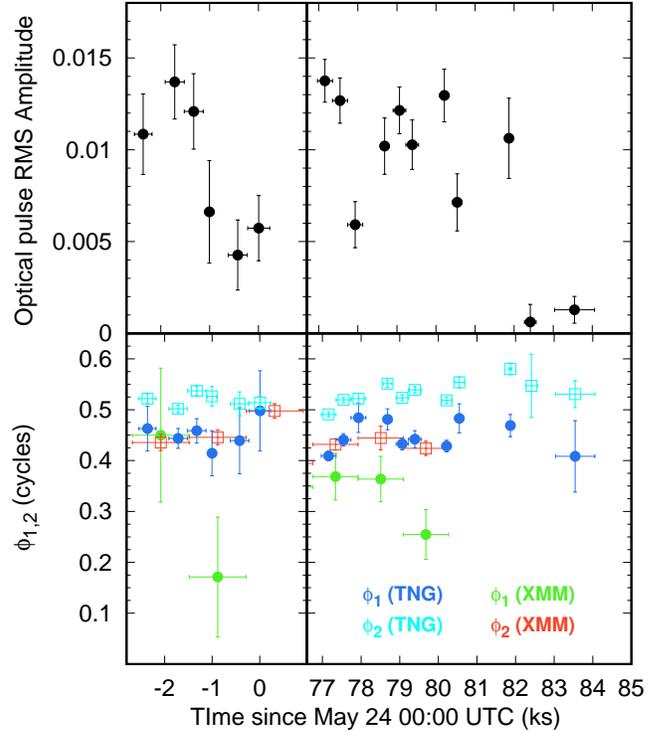}
\caption{ Top panel: the RMS amplitude of the optical pulse along the
  May 23 (left) and May 24 (right) observing runs during the high
  mode.  Lower Panel: the phases of the optical pulse as measured in
  the May 23 (left) and May 24 (right) TNG runs (blue dots and sky blue
  empty squares were used for the first and second harmonic,
  respectively), and the phases of the X-ray pulse as measured in the
  two XMM-Newton runs (green dots and red empty squares were
  used for the first and second harmonic,respectively).}
 \label{fig:phases}
 \end{figure}

We used the {\it low}, {\it high} and {\it flaring} mode time
intervals determined using the EPIC light curves to select the
corresponding  intervals during the TNG
observations. Table~\ref{tab:periods} lists the properties of the
optical pulse during the three modes. We detected optical pulses at a
high confidence level during the {\it high} mode (red points in
Fig.~\ref{fig:chicurves} show the periodogram) with an average,
background-subtracted RMS amplitude of $(0.68\pm0.02)\%$.
Note that since the background is probably overestimated during TNG1
(see Sec.~\ref{sec:tngobs}),  the actual intrinsic pulse
  amplitude was likely slightly larger.  The top panel of
Fig.~\ref{fig:phases} shows the evolution of the RMS amplitude over
{\it high} mode time intervals of length ranging from $\sim200$ to
$450$~s; it attains values as high as $R_{max}=(1.4\pm0.1)\%$, and is
variable although with no clear correlation with the orbital
phase. Towards the end of TNG2 the pulse amplitude decreased down to a
level comparable to that observed during flares ($<0.2\%$, see below),
suggesting that the entire interval from $\sim82$~ks since May 24
00:00 to the end of TNG2 (see right panels of Fig.~\ref{fig:lc2}) was
in the {\it flaring} mode. By considering only the {\it high} mode
intervals before the onset of such a long {\it flaring} event, the
average pulse fraction observed in the {\it high} mode increased to
 $R_{HS}=(0.97\pm0.02)\%$.  Similar to the X-ray band, the optical
pulse was not detected during the {\it low} mode down to an upper
limit of $0.02\%$ (95\% confidence level), i.e roughly 30 times
smaller than the amplitude observed during the {\it high} mode. The
{\it flaring} mode took place only during TNG2 and the optical pulse
was detected at a significance of 8.3$\sigma$ and with amplitude
$R_{2,flare}=(0.16\pm0.02)\%$, i.e. more than five times smaller than
during the {\it high} mode. Considering that the net count rate
observed by SiFAP/{\tng} during flares is about twice that in the {\it
  high} mode, the amplitude decrease is larger than what would be
expected if the flares were a simple addition of unpulsed flux  to
  the {\it high} mode level.  We checked that optical pulses were
detected at a significance larger than 4.5$\sigma$ even when flares
were identified by using a higher threshold in the EPIC X-ray light
curve, $20$~s$^{-1}$, rather than $11$~s$^{-1}$, the value used by
\citet{bogdanov2015} and throughout this paper.

We performed pulse phase timing on the optical pulse profiles computed
over intervals of length spanning between 200 and 500~s in the {\it
  high} mode. We measured an optical period
$P_{opt}=1.687987445(1)$~ms, a value compatible with $P_{ref}$ and
$P_X$, and the epoch of passage at the ascending node
$T^*_{opt}=57896.829267(6)$~MJD, compatible with the X-ray estimate
within $\delta T^* \sim 0.5$~s. This estimate can be used to constrain
the position of the region emitting the optical pulses within $a_1
\delta T^* / P_{orb} \simeq 3 (\sin(i))^{-1} $~km in azimuthal
distance along the orbit from the region emitting the X-ray pulses,
where $a_1$ is the semi-major axis of the pulsar orbit and $i$ is the
system inclination.

Red and green points in Fig.~\ref{fig:pulse} show the normalized and
background-subtracted optical pulse profiles observed in the {\it
  high} and {\it flaring} mode, respectively. The optical pulse is
described by two harmonics with an amplitude ratio $r_2/r_1\simeq1$,
smaller than that of the X-ray pulse ($\simeq3$; blue points in
Fig.~\ref{fig:pulse})). The phase of both harmonic components of the
optical pulse profile lag the corresponding components of the X-ray
profile; the lags of the first and second harmonic are
$\delta\phi_{1}=0.14\pm0.01$ and $\delta\phi_{2}=0.112\pm0.004$,
respectively.  The phase difference observed during periods of
strictly simultaneous observations is compatible with these estimates,
indicating that the phase lag is not due to variability of the X-ray
pulse profile in intervals not overlapping with the optical
observations. The bottom panel of Fig.~\ref{fig:phases} shows the
evolution of the phase of the two harmonic components of the optical
(blue dots and  cyan hollow squares for the first and
second harmonic, respectively) and X-ray (green dots and red hollow squares) pulse profiles over the interval of
simultaneous coverage; a phase shift of $\approx0.1$ is always
observed, with a somewhat larger timing noise shown by the optical
pulse.

 We measured the average optical flux $F_{abs}$ in different modes by
 scaling the observed background-subtracted count rate by the
 conversion factor determined in Sec.~\ref{sec:optflux}.  We estimated
 the pulsed luminosity by scaling these flux values for the ratio
 between the de-reddened and the absorbed flux in the 320--900~nm
 energy band ($k_{der}=1.217$) and multiplying for the total RMS
 amplitude we obtained the pulsed luminosity shown in the rightmost
 column of Table~\ref{tab:periods}. The values obtained in this way
 are listed in of Table~\ref{tab:periods}.  A red point in
 Fig.~\ref{fig:spectrum} shows the average de-reddened pulsed flux
 observed during the high mode
 ($5.2(4)\times10^{-14}$~erg~cm$^{-2}$~s$^{-1}$) in $\nu F_{\nu}$
 units, computed assuming a constant value over the 320--900~nm band.

 \subsubsection{December 2017 campaign. The optical pulse}

\label{sec:dec2017}

\begin{figure}[t!]
  \includegraphics{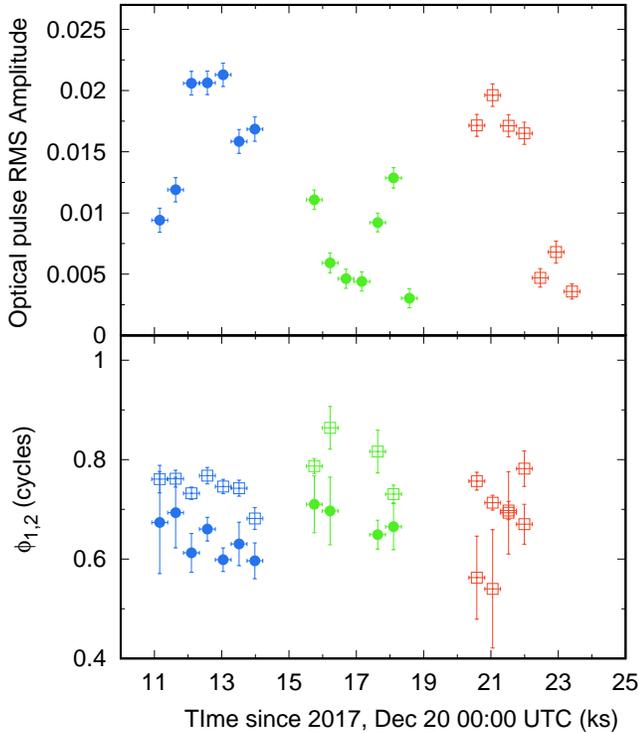}
   \caption{The top panel shows the RMS amplitude of the optical pulse
     profiles observed during TNG3-B (blue points), TNG3-V (green
     points) and TNG3-R (red points) by folding the time series around
     $P_{ref}$ and choosing MJD~$58107.0$ as the reference epoch. The
     bottom panel shows the phase computed over the first (filled
     circles) and the second harmonic (hollow squares). }
 \label{fig:colors}
 \end{figure}

We preliminary measured the epoch of passage of the pulsar at the
ascending node correcting the time of arrival of TNG3 with the
semi-major axis and orbital period of the J16 timing solution, and
varying a grid of values of $T_{asc}$ spaced by 0.2~s around the
expected value. We corrected the event arrival times of TNG3 with the
preliminary value of $T_{asc}$, and folded the resulting time series
around $P_{ref}$. We measured the pulse phase of the first and second
harmonic of the pulse and determined $P_{TNG3}=$1.687987456(12)~ms
(compatible with the value expected according to the J16 solution,
1.687987446177(4)~ms), and $T^*_{TNG3}=58107.009511(4)$~MJD. Blue,
green and red points in Fig.~\ref{fig:colors} mark the RMS amplitude
(top panel) and the phases (dots show those computed on the first
harmonic, hollow squares indicate phases of the second harmonic) of
the pulse profiles observed during the exposures performed with B, V
and R filters, respectively, and measured using the J16
  ephemerides and the same reference epoch used fold May 2017
  data. The phase difference of $\simeq 0.2$ cycles shown by the
  phases computed over the first and second harmonic with respect to
  the values observed during May 2017 (see Table~\ref{tab:periods} and
  bottom panel of Fig.~\ref{fig:phases}) is compatible with the $0.1$
  cycles phase uncertainty obtained propagating the errors on the
  ephemerides given by \citealt{jaodand2016} to the epoch of TNG3. A
  comparison between the pulse phase measured at different epochs is
  further hampered by the noise affecting the measured epoch of
  passage at the ascending node (see Fig.~\ref{fig:tstar}). Note that
the absence of simultaneous X-ray observations prevented us from
identifying transitions between the {\it high} and {\it low} modes
(flares are not evident in the light curves); since pulses are absent
in the {\it low} mode, the amplitudes plotted in Fig.~\ref{fig:colors}
are underestimated with respect to the values measured in the {\it
  high} mode alone. Note that the source spends on average $\sim
70-80\%$ of the time in the {\it high} mode (J16).

 \subsubsection{December 2017 campaign. The X-ray pulse}

\begin{figure}[t!]
  \includegraphics{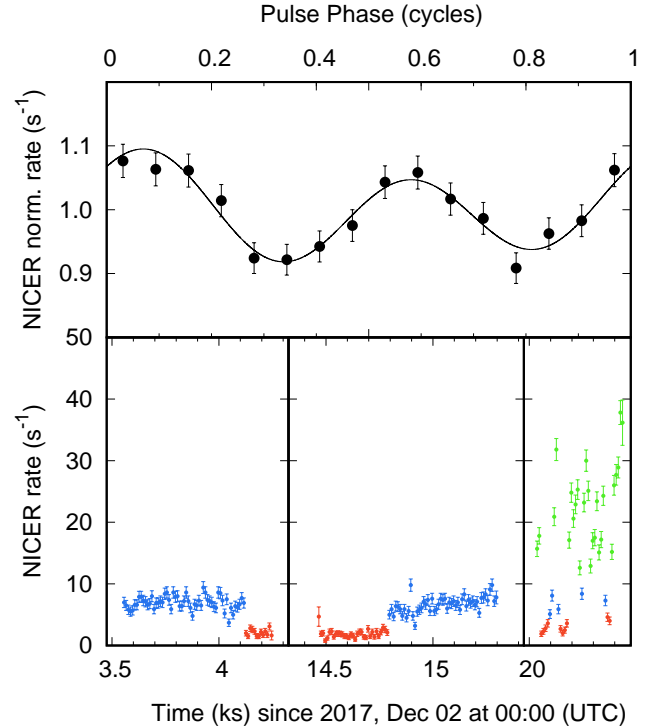}
  \caption{Pulse profile obtained by folding the {\it high} mode
      intervals in the {\nicer} data with respect to the reference
      epoch MJD 57896.0, using J16 ephemerides and the epoch of
      passage at the ascending node determined from December 2017
      optical data (see Sec.~\ref{sec:dec2017}),
      $T^*_{TNG3}=58107.009511(4)$~MJD (top panel). Light curve
      observed by {\nicer} during observation 1034060101 (see
      Tab.~\ref{tab:log}, bottom panel).}
 \label{fig:nicer}
 \end{figure}

 We analyzed three {\nicer} observations performed in December
  2017. The bottom panel of Fig.~\ref{fig:nicer} shows a sample light
  curve observed during the first observation, where the high, low and
  flaring modes can be easily recognized.
  
 Adopting the orbital ephemerides from the timing analysis of the TNG3
 data, we corrected the {\it NICER} photon arrival times and searched
 for coherent pulsations by computing power density spectra over
 2~ks-long intervals. A statistically significant excess at
 $\nu=1184.8426(5)$~Hz is observed in the average power density
 spectrum, consistent with the second harmonic of the spin frequency
 of the source. We applied epoch-folding search techniques to the
 available {\it NICER} observations around the spin period
 extrapolated from the coherent signal detected with the Fourier
 analysis. We obtained the best profile for the spin period
 $P_{NICER}=$1.6879874457(4)~ms, compatible with the value expected
 according to the J16 solution.  The average profile is characterized
 by an RMS amplitude $R=(5.3\pm0.7)\%$.  Extrapolating the count rate
 threshold adopted for {\xmm}/EPIC-pn data, we investigated coherent
 signals in the time series restricted to the three modes. Similarly
 to the {\xmm}/EPIC-pn case, we did not detect any signal in the {\it
   low} and {\it flaring} modes. On the other hand, we detected
 pulsation in the {\it high} mode with amplitude of $(5.6\pm0.9)\%$ ,
 with a second harmonic roughly three times times stronger than the
 fundamental (see top panel of Fig.~\ref{fig:nicer}).
 
 \subsection{The optical spectrum}
\label{sec:optsp}
 
\begin{figure}[t!]
 \includegraphics{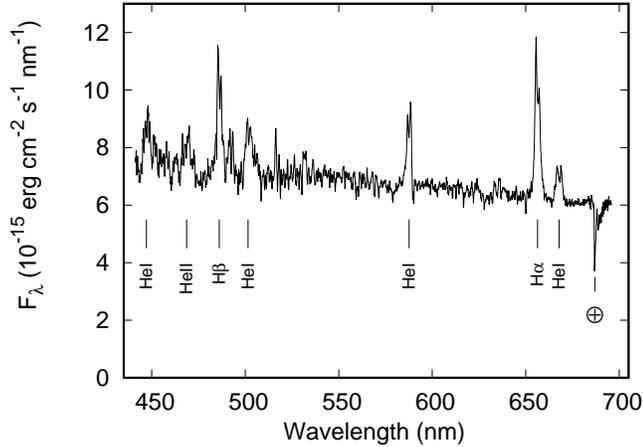}
   \caption{Optical (440--695~nm) spectrum of {\1023} taken with
     ALFOSC at the NOT. The most prominent emission lines are
       marked with a vertical segment. The spectrum was
     normalized to match the flux density observed simultaneously by
     TJO in the V-band ($V=16.85\pm0.02$~mag, corresponding to
     $(6.9\pm0.1)\times10^{-15}$~erg~cm$^{-2}$~s$^{-1}$~nm$^{-1}$ at
     550~nm. During the same interval of NOT observations, the OM
     on-board {\xmm} observed a count-rate of $6.1(1)$~s$^{-1}$, which
     translates into a flux density of
     $7.9\times10^{-15}$~erg~cm$^{-2}$~s$^{-1}$~nm$^{-1}$ at 450~nm. }
 \label{fig:optspectrum} 
 \end{figure}

Figure~\ref{fig:optspectrum} shows the optical spectrum observed by
ALFOSC/{\it NOT}, calibrated in flux using the simultaneous
measurement of the magnitude by the TJO (see
Sec.~\ref{sec:tjo}). Broad, in a few cases double peaked emission
lines of H$\alpha$ 656.3, H$\beta$ 486.1, HeI 587.6, HeI 667.8, HeI 501.5,
HeII 468.6, HeI 447.1~nm are most prominent. Telluric contamination produced
       the absorption line at $\simeq 687$~nm. These features have
double-peaked profiles which are signature of an accretion disk viewed
at moderate inclination.  As the continuum is blue and smooth
\citep[see Fig.~2][]{wang2009}, we created a template spectral shape,
$f_{1023,NOT}(\lambda)$ by extrapolating the continuum observed at
      {\it NOT} to the 320--900~nm band, giving a flux of
      $F_{J1023,NOT}\simeq3.8\times10^{-12}$~{\flux} over this band.

 \section{Discussion}

 This paper presented the first high time resolution
   optical/X-ray/IR/UV observational campaign of {\1023} in the disk
   state.  Similar to other transitional millisecond pulsars in such
 a state, the X-ray light curve of {\1023} shows three intensity modes
 (dubbed {\it low}, {\it high} and {\it flaring}), and coherent X-ray
 pulsations were detected only in the {\it high} mode; they have
 an rms amplitude of $\simeq8\%$ and are detected up to 45~keV,
 as  first shown in this study based on {\nustar} observations.

 Our simultaneous optical/X-ray observations revealed that  also
 optical pulses are observed in the {\it high} mode. Optical
   pulses have an average  rms amplitude of
 $R_{HS}=(0.97\pm0.02)\%$, corresponding to a pulsed luminosity of
 $\bar{L}_{HS}=1.11(5)\times10^{31}$~erg~s$^{-1}$ (see
 Table~\ref{tab:periods}). The spin period and the epoch of passage at
 the ascending node measured from the optical pulses are consistent
 with the values measured in the X-ray band within
 $(0.6\pm1.2)\times10^{-12}$~s and $(0.4\pm0.5)$~s, respectively. The
 latter estimate indicates an azimuthal distance of a few km at most
 between the region emitting the optical and X-ray pulses,
   indicating that the optical and X-ray pulses are produced in the
   same region. Assuming a constant flux over the 320--900~nm band,
 $F_{\nu}\sim$~count's, we estimated the average pulsed flux density
 of $\bar{S}_{\nu}= 10.7(6)\,{\mu}$Jy. This value is compatible with
 the low-frequency extrapolation of the $F_{\nu}\propto \nu^{-0.7}$
 trend that holds for the pulsed flux density in the X-ray band (see
 Fig.~\ref{fig:spectrum}), suggesting that such a component could
 describe the pulsed spectral energy distribution over at least four
 decades in energy.  The optical pulse amplitude  was highly
   variable between 0.5 and 1.5\% over 500~s-long intervals, 
   corresponding to a maximum pulsed luminosity of
 $L_{HS}^{max}\simeq1.6(1)\times10^{31}$~erg~s$^{-1}$.

Similar to X-ray pulses, optical pulses were not detected in the {\it
  low} mode, with an upper limit of $R_{LS}<0.034\%$ (95\% confidence
level). This corresponds to a pulsed luminosity of
$4\times10^{29}$~erg~s$^{-1}$, i.e. more than 25 times smaller than
the value observed in the {\it high} mode.

The simultaneous detection of pulses in the optical and X-ray bands
breaks down during flares. Optical pulses were detected with an rms amplitude of $R_{FL}=(0.16\pm0.02)\%$, corresponding to an
average pulsed luminosity of $0.37(5)\times10^{31}$~erg~s$^{-1}$,
i.e. almost one third than the average value observed in the {\it
  high} mode. X-ray pulsations remained undetected down to an
amplitude of $1.3\%$, corresponding to a pulsed flux $2.3$~times lower
than in the {\it high} mode. If the X-ray pulsed fraction decreases
during flares with respect to the {\it high} mode by the same amount
seen in the optical, we expect them at an amplitude of $\sim 1\%$,
i.e. slightly lower than the upper limit we derived. We conclude that
the non-detection of X-ray pulses during flares may result from
limited photon statistics.

  Both the optical and X-ray pulses were described by the sum of two
  harmonic components, yielding two intensity peaks every NS
  rotation. The ratio of the second to the first harmonic amplitude of
  the optical pulse $r_2/r_1$ was close to unity, lower than the value
  observed in the X-ray band ($\simeq 3$). The first and the second
  harmonic of the optical pulse lagged the X-ray pulse by
  $\delta\tau_1=236\pm17\mu$s and $\delta\tau_2=189\pm7$~$\mu$s,
  respectively. Lags of the same order were observed over intervals of
  few hundreds seconds, in both simultaneous optical/X-ray
  observations analyzed here. The absolute timing accuracy of {\xmm}
  is 48$\,\mu$s \citep{martincarrillo2012}. On the other hand, we
  estimated the {\it SiFAP} absolute time accuracy as better than
  $\sim 60\,\mu$s, by relying on a hour-long observation of the Crab
  pulsar performed at the Cassini Telescope at Loiano Observatory (see
  \ref{app:crab}1). The resulting total systematic error affecting the
  phase lag between the optical and X-ray pulse is $<77\,\mu$s.

Optical pulses were seen at an amplitude varying between 0.5 and 2\%
in the Johnson {\it B}, {\it V} and {\it R} filters. In the absence of
simultaneous observations of a reference star and the X-ray
counterpart we could not measure accurately the optical spectral
energy distribution.

The main result presented in this paper is that optical and X-ray
pulses closely trace the repeated transitions between the X-ray {\it
  high} mode, in which they are both observed and whose pulsed flux is
compatible with a single power law relation $F_{\nu}\propto
\nu^{-0.7}$, and the {\it low} mode, in which they are not
detected. This strongly indicates that optical and X-ray pulses are
related to the same phenomenon, something also suggested by the
similar pulse shape and small phase offset.  In the following we
discuss the implication of these findings for the different scenarios
proposed to explain the enigmatic nature of the disk state shown by
{\1023} and other transitional millisecond pulsars.

\subsection{Accretion onto the NS surface and propelling of matter}

Coherent X-ray pulsations seen in the high mode of {\1023} were 
  first interpreted as due to magnetically channeled accretion onto
the NS hot spots \citep{archibald2015}.  This interpretation was
  justified by the ten-fold increase in the X-ray pulsed and total
flux that occurred after an accretion disk formed in the system in
June 2013, as well as the similarity of the pulsed fraction, shape and
spectrum of the X-ray pulses observed from {\1023} to those shown by
accreting millisecond pulsars. Based on a similar reasoning
\citet{papitto2015} interpreted in terms of accretion power also the
X-ray pulsations shown by the transitional millisecond pulsar {\xss}
in the disk state. However, as both authors noted, interpreting
  such X-ray pulses as due to channeled accretion challenged the
  usual accretion/propeller picture for fast rotating NS. To show
this, we assume that the disk is truncated at a radius equal to $\xi$
times the Alfven radius,

\begin{equation}
  \label{eq:rm}
r_m=57.6~\xi_{0.5}~\mu_{26}^{4/7}\,\dot{m}_{14}^{-2/7}~m_{1.4}^{-1/7}~km
\end{equation}
where $\xi_{0.5}=(\xi/0.5)$, $\dot{m}_{14}$ is the mass accretion rate
in units of $10^{14}$~g~s$^{-1}$ and $\mu_{26}$ is the magnetic dipole
moment in units of $10^{26}$~G~cm$^{-3}$
($\mu_{26}=0.79\,(1+\sin^2\alpha)^{-1}$ for {\1023}, with $\alpha$
angle between the magnetic and the spin axes, evaluated using the
spin-down rate during the radio-pulsar phase measured by J16 into the
relation given by \citealt{spitkovsky2006}).  The observed X-ray
luminosity ($\simeq 8.5\times10^{33}$~erg~s$^{-1}$ in the 0.3--80~keV
band, \citealt{cotizelati2018}) indicates a mass accretion rate of
$\dot{m}_{14}=0.46~m_{1.4}^{-1}$, where $m_{1.4}$ is the NS mass in
units of 1.4~M$_{\odot}$, and according to Eq.~\ref{eq:rm}, a disk
radius of $r_m\simeq72\,\xi_{0.5}$~km. Taking $\xi_{0.5}=1$, this
value exceeds by three times the corotation radius of {\1023}
($r_c$=23.8~$m_{1.4}^{1/3}$~km) and accretion onto the NS surface
should be inhibited by the quick rotation of magnetic field lines at
the magnetospheric boundary.

\citet{papitto2015b} proposed that the magnetosphere would be squeezed
to $\sim r_c$ if the actual accretion rate in the disk were larger
($\dot{m}_{14}=5-17$) than the value indicated by the X-ray
luminosity. In such a case 99\% of the mass should be ejected from the
inner rim of the disk by the propelling magnetosphere in order to
match the relatively low X-ray flux.  Roughly half of the X-ray
emission and the whole gamma-ray output would be due to synchrotron
self-Compton emission by electrons accelerated in the shock formed at
the disk-magnetosphere boundary (see also \citealt{papitto2014}).
Alternatively, the disk could have fallen in a low $\dot{M}$ {\it
  trapped} state \citep{dangelo2012} with its inner rim that stays
close to the corotation radius regardless of how low the mass
accretion rate could get. In the case of {\1023}, such a model would
imply that the observed X-ray luminosity traces the actual mass
accretion rate and a strong outflow would not be launched.

Assuming that X-ray pulses are due to accretion of matter onto
  the magnetic poles, the optical pulses of {\1023} could be explained
  by cyclotron emission by electrons inflowing in magnetized
accretion columns, similar to the case of accreting magnetic white
dwarfs \citep{masters1977,lamb1979}. Indeed, the fundamental
cyclotron energy for electrons inflowing in accretion columns
permeated by a magnetic field of the order of that estimated for
{\1023} \citep[][J16]{archibald2013} is $E_{cyc}=1.2$~eV. Assuming
that $\beta$ is the angle between the magnetic axis and the disk
plane, matter inflowing from a disk truncated at the corotation
radius $r_C$  forms hot-spots on the neutron star surface
of size:

\begin{equation}
A_{spot}\simeq\pi \frac{r_*^3}{r_c} \sin^2{\beta}=1.3\times10^{12}~
\sin^2{\beta}~cm^2,
\end{equation}
 i.e. $\sim 10\%$ of the NS surface, in agreement with the results of
 the simulations performed by \citet{romanova2004}. The corresponding
 typical accretion column transverse length-scale is then
 $\ell\simeq5$~km, and the electron density in the accretion columns for a
 fully ionized plasma is:
 
\begin{equation}
  \label{eq:ne}
n_e= \frac{\mu_e}{m_H} \rho = \frac{\mu_e}{m_H} \frac{\dot{M}}{\pi \ell^2 v_{ff}} = 0.5\times10^{16}\, \dot{m}_{14}\, \ell_5\, m_{1.4}^{-1/2}\, cm^{-3}
\end{equation}
where $\mu_e=1.18$ is the mean molecular weight per electron, $m_H$ is
the proton mass, $v_{ff}=\sqrt{2GM/r_*}$ is the free-fall velocity
close to the NS surface and $\ell_5\equiv\ell/5~\mbox{km}$. The
resulting optical depth to cyclotron absorption of accretion columns
filled by plasma of such density and permeated by a $B\sim 10^8$~G
magnetic field is as large as $\approx 10^6$ in the transverse
direction \citep{trubnikov1958}. This ensures that  the emission
  of the first few cyclotron harmonics is self-aborbed (up to roughly
  ten), and that the resulting spectrum is described by the
Rayleigh-Jeans section of a black-body spectrum with temperature equal
to the electron temperature, $T_e$.

However, in \citet{ambrosino2017} we estimated the maximum 
  cyclotron luminosity expected in the 320-900~nm band as:
\begin{eqnarray}
  \label{eq:cyclo}
L_{cyc}&=&A_{spot} \int_{\nu_l}^{\nu_h} (2 \pi k T_e \nu^2 /  c^2) d\nu = \nonumber \\&=& 2.9\times 10^{29} \left(\frac{A_{spot}}{10^{12}cm^2}\right) \left(\frac{kT_e}{100 keV} \right)\,erg\,s^{-1},
\end{eqnarray}
where $\nu_l=3.33\times10^{14}$~Hz and $\nu_h=9.37\times10^{14}$~H< are the boundaries of the band observed by
{\it SiFAP}.  This value is $\approx 40$ times lower than the
observed average  optical pulsed luminosity. This
  discrepancy holds even when taking for $kT_e$ a value of $\sim
100$~keV, of the order of that observed from accreting millisecond
pulsars \citep[][and references therein]{patruno2012}, and likely an
overestimate of the temperature attained by electrons in the accretion
column of a pulsar with an accretion luminosity of
$<10^{34}$~erg~s$^{-1}$. In fact, such a high  electron
  temperature can be reached if the pressure exerted by the radiation
emitted from the hot-spots balances the gravitational inflow of plasma
in the accretion columns, and forms a shock standing off the NS
surface, where the kinetic energy of the flow is converted into
thermal motion of the charges. The {\it critical} luminosity to form
such a shock is $L_{crit}\approx 10^{36}\,\ell_5$~erg~s$^{-1}$
\citep{basko1976}, more than a hundred times larger than the value
observed from {\1023}. Below such a value the ions of the in-falling
plasma are best slowed down by Coulomb collisions with atmospheric
electrons \citep[see, e.g.,]{frank2002},
and a temperature of the order of the effective black-body temperature
is attained, $kT_e\simeq 1.9 (L/L_{crit})^{1/4}$~keV. We deduce that
magnetic accretion at a rate $L\sim 10^{-3} L_{crit}$ is hardly
capable to produce electrons hot enough to yield a sizable cyclotron
emission in the optical band (e.g. $L_{cyc}\simeq10^{27}$~erg~s$^{-1}$
for $kT_e\simeq0.3$~keV).

According to Eq.~\ref{eq:cyclo}, the maximum observed pulsed optical
luminosity $L_{HS}^{max}\simeq1.6(1)\times10^{31}$~erg~s$^{-1}$ 
  corresponds to an unrealistically large brightness temperature of
$k T_b=(175\pm10)~(r_{em}/\mbox{km})^{-2}$~MeV, where $r_{em}$ is the
radius of the emission region.  Even considering $r_{em}\sim 100$~km
(i.e. approximately the size of the light cylinder), the X-ray
luminosity that would be produced by such a thermal component ($\sim
10^{44}$~erg~s$^{-1}$) would be huge. We can then safely rule out
emission from hot-spots on the NS surface heated by the accretion flow
as the origin of optical pulses.  A similar reasoning rules out
reprocessing of the X-ray emission by the inner regions of the disk,
which would necessarily produce an even cooler and fainter thermal
spectrum,

\subsection{Rotation-powered pulsar}
\label{sec:radiopulsar}
An alternative possibility is that the optical pulsations of {\1023}
originate in the activity of a rotation-powered pulsar. So far,
optical pulsations have been detected from five isolated high-magnetic
field young pulsars \citep[][and references
  therein]{mignani2011}. Models envisage that synchrotron emission of
secondary electron/positron pairs accelerated in magnetospheric gaps,
re-connection events and/or the equatorial current sheet \citep[see,
  e.g.,][for a recent review]{venter2018} give rise to non-thermal
pulsed emission at optical and X-ray energies \citep{pacini1983},
whereas curvature radiation accounts for the gamma-ray emission
\citep{romani1996}. Recently, the need of using a common description
of these processes, dubbed as synchro-curvature, has become evident
since both effects are relevant along the particles trajectories in
the magnetosphere \citep{vigano2015}.  These mechanisms are unlikely
to work if the magnetosphere is engulfed with plasma from the disk
(density of $n_e\simeq 5\times10^{15}$~cm$^{-3}$, see Eq.~\ref{eq:ne},
i.e. $\approx 10^6$ times the Goldreich \& Julian critical density,
\citealt{goldreich1969}) as gaps in the outer magnetosphere would be
readily filled\footnote{Note that \citealt{bednarek2015} proposed a
  coexistence of an equatorial disk flow down to the NS surface and
  electron acceleration in high latitude slot gaps}
\citep{shvartsman1971}. Even if electron acceleration up to a Lorentz
factor $\gamma_3\equiv\gamma/10^3=1$ were possible, charges would be
stopped down by Coulomb collisions with ions and electrons of the
plasma on a typical length-scale of $\approx
5\times10^{-5}\gamma_3^{-1}$~cm \citep[see, e.g., Eq.~3.35
  in][]{frank2002}. This is much smaller than the length over which
electrons radiate synchrotron X-ray and optical photons in a pulsar
magnetosphere, $< 10^{-5}r_{LC}\simeq80$~cm for {\1023} after particle
injection \citep{torres2018}. For this reason we assume that a
rotation-powered pulsar is able to work only if the matter inflow is
truncated outside the light cylinder, and no accretion onto the NS
surface takes place. To meet this condition, the mass accretion rate
should be lower than $\dot{m}_{14}<0.3\,\xi_{0.5}^{7/2}$ (see
Eq.~\ref{eq:rm} for $r_m>r_{LC}$), and the inferred luminosity lower
than $2.5\times10^{32}$~erg~s$^{-1}$. In this scenario most of
the observed X-ray luminosity ($\simeq 8.5\times10^{33}$~erg~s$^{-1}$)
 would not be due to disk accretion.

However, assuming that the optical pulses of {\1023} originate in the
magnetosphere of a rotation-powered pulsar presents a number of
issues. First, a very large efficiency is required to explain the
conversion of up to ${L}_{HS}^{max}/\dot{E}\simeq3.6\times10^{-4}$ of
the spin down power $\dot{E}=4.4\times10^{34}$~erg~s$^{-1}$
\citep{archibald2013} in 320--900~nm optical pulsed luminosity. Values
lower by at least an order of magnitude are observed from other
powerful rotation powered pulsars, including the Crab pulsar
\citep{percival1993} and the isolated millisecond pulsar PSR
J0337+1715 (\citealt{strader2016}; see Fig.~3 of
\citealt{ambrosino2017}, which compares the optical efficiency in the
B-band of various types of pulsars).

Secondly, the pulsed X-ray luminosity in the {\it high} mode was
$\simeq 2.6\times10^{32}$~erg~s$^{-1}$ \citep[][see also
  Table~\ref{tab:periods}]{archibald2015}, i.e. $\sim 6\times10^{-3}$
times the spin down power. The simultaneous detection of optical and
X-ray pulses in the {\it high} mode and their disappearance in the
{\it low} mode means that if the former has a magnetospheric origin,
likely also the latter does. The fraction of the spin-down power
converted in X-ray pulses of {\1023} would be much larger than that of
almost all rotation-powered pulsars ($<10^{-3}$,
\citealt{possenti2002, vink2011}, see also \citealt{lee2018} for an
updated analysis that suggests an average efficiency
$\simeq10^{-4}$). Considering that rotation-powered pulsars with
sinusoidal pulse profiles generally show a pulsed fraction of
$\sim10\%$ \citep{zavlin2007}, the discrepancy is likely even
larger. More importantly, when a disk was absent and radio pulses were
observed, \citet{archibald2010} detected X-ray pulsations in the
0.25-2.5 keV band with an rms amplitude of $(11\pm 2)\%$. Even
assuming that pulsations were present also in the 2.5--10~keV band and
with the same amplitude (note that in that band
\citet{archibald2010} could only place a 20\% upper limit on the pulse
amplitude),  the pulsed X-ray luminosity would have been $\sim
10^{31}$ erg s$^{-1}$, i.e. $2\times 10^{-4}$ times the spin-down
power. The 25-fold increase of the pulsed flux that occurred
when a disk formed in the system would be very difficult to explain
assuming that the rotation-powered pulsar kept working as if it were
in the radio pulsar state. One case of mode switching by an isolated
rotation-powered pulsar is known \citep{hermsen2013,mereghetti2016},
but it appears contrived that the mode change of {\1023} which
occurred when a disk formed was not influenced by it.

The large optical and X-ray spin down conversion efficiency needed to
produce a magnetospheric emission large enough to explain the observed
pulsed flux could be  indeed related to the presence of the disk.
Soft disk photons could enhance the pair production in the
magnetosphere yielding a brighter pulsed radiation than in the radio
pulsar state in which the system roughly behaves as if it were
isolated.  However, a simultaneous fit of the gamma-ray and X-ray
emission of {\1023} with models developed for rotation-powered pulsars
was troublesome. We considered the synchro-curvature model developed
by \citet{torres2018}. The model was shown to be able to describe well
the X-ray and gamma-ray emission of rotation-powered pulsars in terms
of a few order parameters (such as the accelerating electric field, a
measure of how uniform the distribution of particles emitting towards
us, the magnetic gradient along a field line and a
normalization). Particularly, in all cases in which both energy bands
displayed pulsed emission, the spectral model built out of only the
gamma-ray data is close to the detected X-ray emission spectrum
already, and further common analysis of both energy regimes makes for
a perfect agreement \citep{torres2018,li2018}. This has proven not to
be the case here: we attempted to model the gamma-ray/X-ray pulsed
energy distribution observed from {\1023} (see
Fig.~\ref{fig:spectrum}), but even assuming that the gamma-ray
emission comes entirely from the magnetosphere (note that gamma-ray
pulses have not been detected from {\1023} in the disk state, so far),
varying the relevant magnetospheric parameters over the wide range
used in \citet{torres2018} gives an X-ray and optical pulsed output
lower than observed by one and three orders of magnitude,
respectively. Based on the different behavior found for {\1023} when
compared to all other pulsars studied from the synchro-curvature model
we conclude that the magnetospheric activity of a rotation-powered
pulsar that works as if it were isolated (and with most of the
gamma-ray radiation pulsed) is unlikely the only source of the
optical/X-ray pulses observed from {\1023}.

\subsection{Pulsar wind}
\label{sec:pwn}

 \begin{figure*}[t!]
   \includegraphics{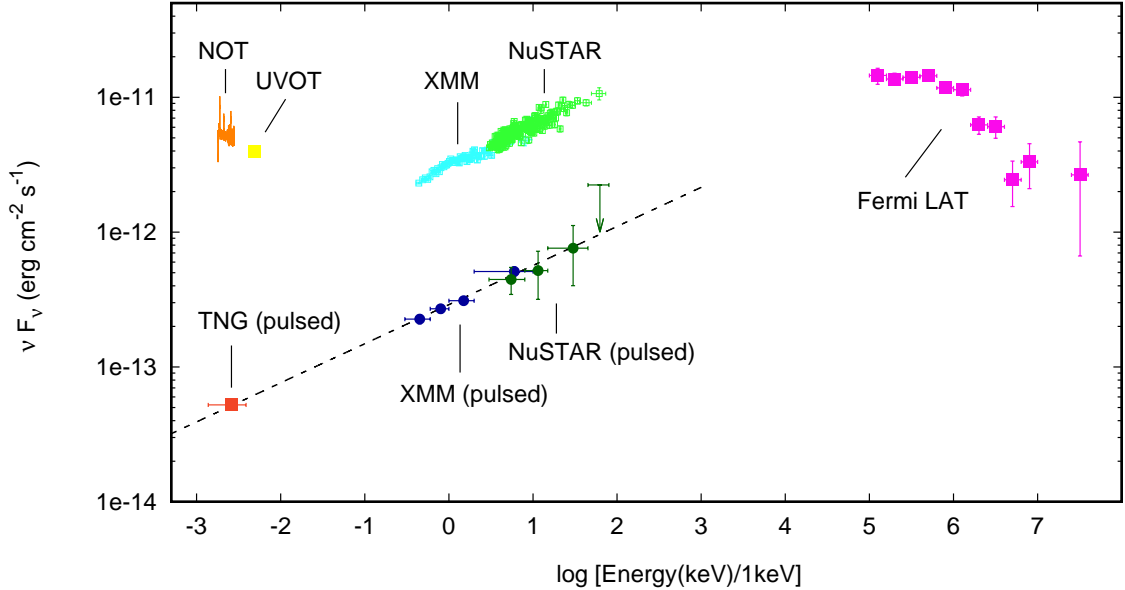}
   \caption{Total and pulsed spectral energy distribution of
     {\1023} in the {\it high} mode, corrected for interstellar
     extinction.  The NOT optical spectrum (see
       Sec.~\ref{sec:optsp}) is plotted with an orange line. The
     average pulsed {\tngsifap} optical flux observed in May 2017 (see
     Sec.~\ref{sec:optmay2017}) is plotted with a red square. The flux
     observed by {\sw}/UVOT UVW1 is plotted with a yellow square.  The
     total and pulsed X-ray flux observed by {\xmm} (see
       Sec.~\ref{sec:xmm2017}) are plotted with light and dark blue
     points, respectively. The total and pulsed X-ray flux observed by
     {\nustar} (see Sec.~\ref{sec:xmm2017}) are plotted with
     light and dark green points, respectively.  Magenta squares show
     the average {\it Fermi}-LAT spectrum measured by
     \citep{torres2017}. The dashed line is a cut-off power law $\nu
     F_{\nu}\sim \nu^{0.3}$ normalized to match the optical and
       X-ray observed pulsed flux.  }
 \label{fig:spectrum} 
 \end{figure*}

\begin{figure}[t!]
   \includegraphics{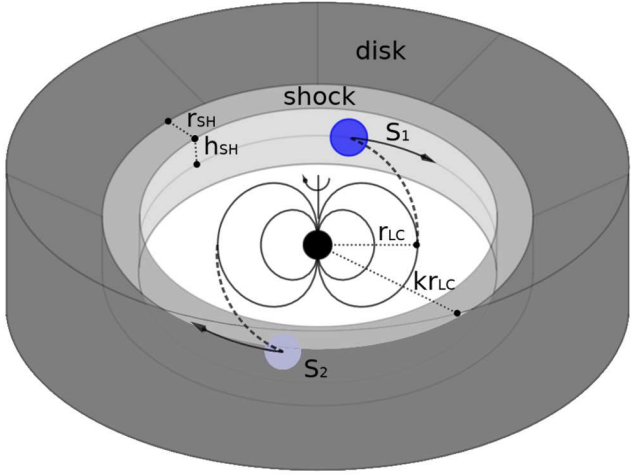}
   \caption{Schematic diagram of the pulsar wind scenario we
       propose to explain the optical and X-ray pulses observed from
       {\1023}. Dashed lines represent the current sheet which expands
       from the light cylinder surface at $r_{LC}$ as an Archimedean
       spiral. Its arms crosses the termination shock (shaded in light
       gray) at S$_1$ and S$_2$ where particles are accelerated to
       relativistic energies. As long as the particle energy is quickly
       radiated away and the size of the shock is smaller than a few
       light cylinder radii, two bright synchrotron-emitting spots
       (drawn in blue) rotate at the shock surface. An observer will
       see pulses of radiation because the intensity received from
       S$_1$ is modulated by the angle under which the spot is seen,
       and the emission coming from S$_2$ is absorbed by the optically
       thick disk in-flow (shaded in dark gray).  }
 \label{fig:model} 
 \end{figure}

We suggest an alternative interpretation of the optical and X-ray
pulses shown by {\1023} in terms of synchrotron radiation emitted from
the intrabinary termination shock of the pulsar wind with the
accretion disk in-flow at a distance $k r_{LC}$, with $k=1$--2,
i.e. just beyond the light cylinder (see Fig.~\ref{fig:model} for
  a schematic diagram of the geometry we have assumed; see also
  \citet{veledina2019} who have presented an interpretation based on
  similar assumptions).  For an isotropic pulsar wind, the post-shock
magnetic field is \citep{arons1993}:

\begin{equation}
\label{eq:bshock}
B_s = 3 \left(\frac{\sigma}{1+\sigma}\right)^{1/2}
\left(\frac{\dot{E}}{c f_{eq} r^2}\right)^{1/2}\simeq 4.5\times10^5\,k^{-1}\,\mbox{G,}
\end{equation}
 where $\sigma$ is the magnetization parameter of the wind
 \citep{kennel1984}, which is $>>1$ close to the light cylinder as the
 whole pulsar wind energy is carried by the electromagnetic Poynting
 flux \citep{arons2002}, and $f_{eq}$ is a geometric factor that
 defines the fraction of the sky into which the pulsar wind is emitted
 and is unity if the wind is isotropic.   For small values of $k$,
   i.e. not far from the light cylinder, the medium is permeated by
 such a large magnetic field that synchrotron emission is the dominant
 cooling mechanism for electrons accelerated at the shock. A single
 population of electrons with energy spectrum $N_E\sim E^{-2.35}$ and
 cut-off at $\sim 2$~GeV \citep[see, e.g., Eq.~35 of][]{lefa2012}
 would produce a spectral energy distribution compatible with the
 shape suggested by the pulsed flux measured both in the optical and
 the 0.3--45~keV X-ray band, $ \nu F_{\nu} \sim \nu^{0.3}$.  (see
 dashed line in Fig.~\ref{fig:spectrum}; \citealt{martin2012}). This
 population could result from a Fermi process with an acceleration
 parameter $\xi\simeq0.01$ (see Eq.~20 in
 \citealt{papitto2014,papitto2015b}).  At low energies the synchrotron
 emission becomes optically thick below $E_{break}=0.9\,
 n_{e,16}^{0.31}\,\ell_{5}^{0.31}\,(B/5\times10^5\,G)^{0.69}$~eV
 \citep{rybicki1979}. Since the electron density at the innermost
 regions of a disk truncated at $Ir_{LC}$ with mass accretion rate
 $\dot{m}_{14}<0.3$ (see Sec.~\ref{sec:radiopulsar}) is $n_e =
 \mu_e/m_p \times \dot{M} / 4 \pi \sqrt{GM} (kr_{LC})^{3/2}\simeq
 5\times10^{12}\,k^{-3/2}$~cm$^{-3}$, the break to optically thick
 emission is expected below $\sim$0.2~eV. This ensures that the
 shock region (see light gray shaded region in
   Fig.~\ref{fig:model}) is optically thin to emission in the visible
 band.

Close to the light cylinder the spin down energy of the pulsar is
transported outwards by the electromagnetic field. In the striped wind
model, the magnetic field configuration has the shape of two monopoles
of opposite polarity which join at the equatorial plane
\citep{bogovalov1999} and a current sheet forms along such a plane
where the field changes polarity. In the oblique rotator case, the
rotation of the pulsar introduces oscillations in the the current
sheet which expands as an Archimedean spiral whose arms are separated
by $\pi r_{LC}=c P_s / 2$ \citep[see Fig.~4
  of][]{bogovalov1999}. Injection of energy at the termination
intrabinary shock then proceeds with a periodicity of $\sim P_s
/2$. The electrons accelerated to relativistic energies at
different locations in the shock will radiate their energy on a time
scale

\begin{equation}
  \label{eq:tsync}
  \begin{split}
t_{sync}&\simeq \frac{\gamma m_e c^2}{P_{sync}} \simeq \frac{9m_e^3 c^5}{4e^4B_s^2\gamma} \\ &\simeq 1.8 \,
\left(\frac{\epsilon}{10~\mbox{keV}} \right)^{-1/2} \, \left(
\frac{B_s}{4.5\times10^5~\mbox{G}} \right)^{-3/2}\,\mu\mbox{s}
\end{split}
\end{equation}
where we used the relations,  $P_{sync}\simeq 4/3 \,\sigma_T c
  \gamma^2 U_B$ for the synchrotron power emitted by a relativistic
  electron and $\epsilon= 0.29 (3 e \hbar / 2 m_e c) B_s \gamma^2 $
\citep{ginzburg1965} to express the energy  of photons at which
  most of the synchrotron power is emitted in terms of the electron
energy $\gamma m_e c^2$.  In the expressions above, $\sigma_T=8\pi
  e^4/ 3 m_e^2 c^4$ is the Thomson scattering cross section and
  $U_B=B^2/8\pi$ is the magnetic energy density. On the other hand,
the light travel time between different regions of the shock is:

\begin{equation}
\label{eq:tlight}
t_{lt}=2 k r_{LC} \sin{i} / c = k P_S \sin{i} / \pi = 533\, k \sin{i} \,\mu\mbox{s}.
\end{equation}
 As long as both the emission time scale and the light travel time
 between different locations of the emission regions are shorter than,
 say, half the spin period, a synchrotron emitting spot (shaded
   in blue in Fig.~\ref{fig:model}) will be seen to rotate coherently
 at the shock surface with a period $P_S/2$. For a moderately
   large inclination angle, the exact value depending on the disk and
   shock relative height, the disk will absorb the emission coming
   from the spot closest to the observer (labeled as $S_2$ in
   Fig.~\ref{fig:model}). On the other hand, the emission from the
   spot located farther from the observer ($S_1$) will be modulated
 sinusoidally as the spot rotates, as if the wind-disk shock were a
 sort of reflecting mirror. Two pulses of optical/Xray synchrotron
   radiation will be then observed every spin cycle of the pulsar,
   whose relative amplitude depends on the magnetic inclination angle,
   as well as on the viewing angle.  Relativistic beaming and/or an
 ordered magnetic field would increase the anisotropy of the emitted
 radiation and the pulse amplitude. In this scenario, the large duty
 cycle of the observed X-ray pulse could result from the sum of the
 periodic emission emitted from different locations of the intrabinary
 shock, which are reached at the different times by the spiralling-out
 current sheet, and are seen at different angles from the observer.
 Interestingly, the synchrotron timescale for optical photons is $\sim
 180\,\mu$s, compatible with the lag of optical pulses with respect to
 X-rays. The observed lag would find an immediate interpretation in
 our modeling, as optical synchrotron photons take a longer time to be
 emitted than X-rays.

The synchrotron timescale expressed by Eq.~\ref{eq:tsync} increases if
the shock is located at a greater distance  because the strength
  of the post-shock magnetic field decreases linearly with distance
(see Eq.~\ref{eq:bshock}). Eventually, it becomes comparable to half
the spin period for 1~eV photons when the magnetic field  at the
  shock is as low as $1.5\times10^5$~G. To produce optical coherent
oscillations the shock surface must be then located at $k\simlt 2$
times the light cylinder radius. The condition of the light travel
time of different regions of the shock, $\approx 2 k r_{LC}
\sin{i}/c=k P_s / 2\pi<P_s/2$, implies a similar constraint, $k<\pi/(2
\sin{i})$. Remarkably, the latter condition is geometrical and does
not depend on the energy of the photons. We speculate that the
simultaneous disappearance of X-ray and optical pulses during the {\it
  low} mode might be due to the inner rings of the mass inflow being
pushed outwards by the pulsar wind, corresponding to an expansion of
the termination radius beyond $\simeq \pi/(2 \sin{i})$ ($k \simeq 2.2$
for $i=45^{\circ}$) times the light cylinder radius. This is in
agreement with the observation of radio flares during the X-ray {\it
  low} modes by \citet{bogdanov2018}, who interpreted them as episodes
of ejection of optically thin plasmoids by the rotation-powered
pulsar.

The synchrotron emission timescale expressed by Eq.~\ref{eq:tsync} is
shorter than the flight time of accelerated particles in the shock
region, $r_{SH}/c \simeq 33\, (r_{SH}/10\mbox{km})\, \mu$s,
provided that the latter is larger than a few km. This ensures that
the energy of the electrons is radiated away before they escape from
the acceleration region.

The energy radiated in pulsed X-rays is $5\times10^{-3}$ times the
spin down energy. The X-ray efficiency of typical isolated pulsar wind
nebulae is usually of a few per cent of the spin down power
\citep{kargaltsev2010}, and rarely reaches a value larger than ten per
cent out of the reverberation phase
\citep{younes2016,torres2018b}. Assuming an efficiency of the order of
that observed from the Crab pulsar nebula (0.04), roughly $10\%$ of
the pulsar wind energy must be converted into X-rays to match the
pulsed flux observed from {\1023}. For an isotropic distribution of
the pulsar wind, the shock height must be $h_{SH} \sim 0.2 kr_{LC}$ to
meet the energy requirement. This is ten times larger than the height
a Keplerian disk would have at a similar distance for a mass accretion
rate $\dot{m}_{14}$, indicating that the shock must be vertically
  extended.  An even more extended shock is required if the whole
  X-ray luminosity seen in the {\it high} mode ($\simeq 0.2\dot{E}$)
  is due to synchrotron emission in the intrabinary shock, suggesting
  a higher efficiency of conversion of the pulsar wind in electron
  energy than previously assumed. A detailed modeling of the
  multi-wavelength spectral energy distribution of {\1023}, including
  also the component observed at GeV energies will be presented in a
  forthcoming paper.

In the context of a variable shock height $h_{SH}$, an increase of the
solid angle covered by the shock could explain flares observed
simultaneously in the optical and X-ray bands. As these sometimes
reach a luminosity comparable to the spin down power of {\1023}
\citep{bogdanov2015}, it is evident that almost complete enshrouding
of the pulsar by disk plasma and a conversion efficiency of spin down
power into electron energy close to unity would be required. It is
unclear, however, why X-ray pulsations should disappear during
flares. As we pointed out earlier, the non-detection could be due to
the lower counting statistics in the X-ray band than in the optical
band. The pulsed optical luminosity observed during flares is roughly
a third of that observed in the {\it high} mode; a smaller
  decrease of the amplitude would be expected if flares were a sheer
superimposition of unpulsed flux over the {\it high} mode level. In
addition, we observed a similar decrease in the optical pulse
amplitude in intervals that formally fell in the {\it high} mode
according to our definition, but were observed in-between flares
(i.e. towards the end of TNG2, see Fig.~\ref{fig:lc2}). This would
suggest that flaring intervals unlikely result from the addition of a
component to the {\it high} mode emission, and should be treated
separately.  However, it is also possible that flares are produced
  in the outer regions of the disk, and the marked decrease of the
  optical pulse amplitude during flares could be due to the occurrence
  of {\it low modes} that cannot be identified from the X-ray light
  curve as they are out-shined by the {\it flaring} emission. More
  observations of the pulsed amplitude decrease during flares are
  needed to break the degeneracy and identify the region where flares
  are produced.

 The possibility that the accretion flow is stopped by the pulsar
wind just beyond the light cylinder by the pulsar wind was recently
explored with general relativistic MHD simulations by \citet[][see
  panel d in their Fig.~4]{parfrey2017}, who noted that X-ray emission
would be expected from the trains of shocks and sound waves produced
at the pulsar wind termination.  They found that in this scenario the
amount of open magnetic flux was similar to the isolated pulsar
case. The spin down rate of the pulsar was then expected to be similar
to that observed in the rotation-powered state. This is qualitatively
in agreement with the $\sim30\%$ increase in the spin down rate
observed by J16 after the formation of an accretion disk, a factor
much smaller than that expected if accretion and/or propeller ejection
took place.

\citet{eksi2005} showed that a stable equilibrium between the
  outward pressure of a rotation-powered pulsar and the inward
  pressure of the infalling matter can be realized if the termination
  shock is close to the light cylinder; this is due to the presence
of a transition region from the near region inside the light cylinder,
where the energy density of the electromagnetic field scales as $\sim
r^{-6}$, to the radiation zone far outside the light cylinder, where
the $\sim r^{-2}$ scaling holds. Equilibrium solutions for values of
the disk truncation radius $k r_{LC}$ with $k\simgt 1$ were found, the
exact value depending on the angle between the magnetic moment and the
rotation axis, $1-\alpha$. For instance, they found stable solutions
with $1<k<2$ for $\xi=10^{\circ}$ and $\mu_{26}=1$, compatible with
the assumptions of our model. On the other hand, far from the
  light cylinder the radiation of a rotation-powered pulsar ($\sim
r^{-2}$) decreases less steeply than the ram pressure of matter
in-falling under the gravitational pull of the compact object ($\sim
r^{-5/2}$). Because of this, a stable solution of a
rotation-powered pulsar surrounded by an accretion disk would not
ensue as the disk is expected to be fully ablated away by the pulsar
wind as soon as $r_m>>r_{LC}$ \citep{shvartsman1970,burderi2001}.
\citet{takata2014} and \citet{CZ2014} modeled the multi-wavelength
emission of {\1023} assuming that the rotation-powered pulsar wind
interacted with the accretion disk to produce synchrotron X-ray
emission far from the light cylinder ($\sim 10^9$~cm, i.e. $k\sim
125$). Besides the problems of stability that would probably
  ensue, we note that in the framework we propose coherent pulsations
would not be produced at such a large distance from the pulsar.

Assuming that most of the pulsed emission is produced at the
intrabinary shock between the pulsar wind and the disk does not rule
out that magnetospheric rotation-powered pulses are still produced at
a similar level than observed during the radio-pulsar state. However,
even if a radio pulsar were active, its pulses would be smeared and
absorbed by material ejected by the pulsar wind
\citep{stappers2014}. On the other hand, the five-fold increase in
gamma-ray flux observed after the disk formation could be due to
Compton up-scattering of disk UV photons off the cold relativistic
pulsar wind, as proposed by \citet{takata2014}, while the pulsed
magnetospheric emission keeps working at a similar rate as in the
radio pulsar state \citep{tam2010}.

\section{Conclusions}

We presented the first simultaneous optical and X-ray high temporal
resolution observations of {\1023} in the accretion disk state, the
only optical millisecond pulsar discovered so far
\citep{ambrosino2017}. We showed that optical pulsations are detected
during the {\it high} mode observed in the X-ray light curve, in which
also X-ray pulsations appear, while pulsations in both bands disappear
in the {\it low} mode. Optical pulses are described by two harmonics,
similar to X-ray pulses and lag the X-ray pulses by $\sim 200\,\mu$s,
although we caution that the absolute time calibration of {\it SiFAP}
is based on just a single observations (see \ref{app:crab}1).  These
findings suggest that the same phenomenon produces the pulses seen in
both energy bands. Cyclotron emission from matter accreting onto the
polar caps of the NS is not powerful enough to explain the pulsed
optical luminosity. On the other hand, emission from the magnetosphere
of a rotation-powered pulsar requires an unusually large efficiency of
spin down power conversion to match the optical and X-ray pulsed flux
of {\1023} with respect to other known pulsars. We argued that {\1023}
is a rotation-powered pulsar whose relativistic, highly magnetized
wind interacts with the inflowing disk matter just beyond the light
cylinder, creating a shock where the wind periodically deposits energy
by accelerating electrons at the shock; these in turn produce optical
and X-ray pulses through synchrotron emission. This would make {\1023}
the prototype of a few hundred km-sized pulsar wind nebula, and
provide an unique opportunity to study the pulsar wind properties in
the high magnetization regime rather than where they are
particle-dominated, as in the usual sub-parsec scale pulsar wind
nebulae. This scenario also provides an explanation of the {\it low}
mode and {\it flares} observed in the X-ray light curve in terms of
the shock being pushed back by the pulsar radiation or increasing its
size, respectively.  Future observations will confirm the phase lag
between optical and X-ray pulses, study the energy distribution of the
pulses in the visible band, search for polarized pulsed emission, and
look for gamma-ray pulsations, thus testing the scenario we
proposed. On the other hand, magneto-hydrodynamic simulations will be
performed to demonstrate that pulsed emission can be indeed generated
by a disk/wind intrabinary shock close to the light cylinder of a
pulsar. The stability of a wind-disk shock just beyond the light
cylinder over timescales of years should also be investigated.  Other
transitional millisecond pulsars in the sub-luminous disk state such
as {\xss} \citep{demartino2010,demartino2013,bassa2014} and candidates
like 1RXS J154439.4--112820 \citep{bogdanovhalpern2015} ans 1SXPS
J042947.1--670320 \citep{strader2016}, XMM J083850.4--282759
\citep{rea2017} and CXOU J110926.4-650224 \citep{cotizelati2019} may
be found in a similar state to that {\1023}. A search of optical
pulsations in those source seems therefore warranted.

\acknowledgments{We are grateful to the {\xmm}, {\tng} and {\it GTC}
  directors for scheduling ToO observations in the Director
  Discretionary Time, as well as to all the teams involved for the
  effort in scheduling the simultaneous ToO observations of {\1023}.
  This work is partly based on observations made with the Italian
  Telescopio Nazionale {\it Galileo} (TNG), operated on the island of
  La Palma by the Fundaci{\'o}n Galileo Galilei of the Istituto
  Nazionale di Astrofisica (INAF), and with the Gran Telescopio
  Canarias (GTC), which are installed in the Spanish Observatorio del
  Roque de los Muchachos of the Instituto de Astrofísica de Canarias,
  in the island of La Palma.  Some of the scientific results reported
  in this study are based on observations obtained with {\xmm}, which
  is a European Space Agency (ESA) science mission with instruments
  and contributions directly funded by ESA Member States and NASA,
  with the {\swfirst}, which is a NASA/UK/ASI mission, with the
  {\nustar} mission, which is a project led by the California
  Institute of Technology, managed by the Jet Propulsion Laboratory
  and funded by NASA, with {\nicer}, which is a NASA mission.
  Development of CIRCE was supported by the University of Florida and
  the National Science Foundation (grant AST-0352664), in
  collaboration with IUCAA.  AP acknowledges funding from the EUs
  Horizon 2020 Framework Programme for Research and Innovation under
  the Marie Skodowska-Curie Individual Fellowship grant agreement
  660657-TMSP-H2020-MSCA-IF-2014. We acknowledge financial support
  from the Italian Space Agency and National Institute for
  Astrophysics, ASI/INAF, under agreements ASI-INAF I/037/12/0 and
  ASI-INAF n.2017-14-H.0. DdM also acknowledges support from PRIN-INAF
  SKA/CTA Presidential decree N. 70/2016. AP, FCZ and DT acknowledhe
  the International Space Science Institute (ISSI-Beijing) which
  funded and hosted the international team "Understanding and unifying
  the gamma rays emitting scenarios in high mass and low mass X-ray
  binaries". AP acknowledges the Cost Action PHAROS (CA16214) which
  funded the international workshop "The challenge of transitional
  millisecond pulsars", hosted by the INAF Osservatorio Astronomico di
  Roma.  NR is supported by the ERC Consolidator Grant "MAGNESIA"
  (nr. 817661), DFT, NR, FCZ, are support by the Spanish Grant
  PGC2018-095512-B-I00, the Catalan Grant SGR2017-1383, and the COST
  Action "PHAROS" (CA 16124). AV was supported by the grant
  14.W03.31.0021 of the Ministry of Science and Higher Education of
  the Russian Federation. SSE was supported in part by a University of
  Florida Research Foundation Professorship. YD was supported in part
  by a University of Florida Graduate Student Fellowship. AG
  acknowledges L.~Di Fabrizio and L.~Riverol for the manufacturing of
  the original mask used in SiFAP. FO acknowledge the support of the
  H2020 Hemera program, grant agreement No 730970.

}
  
    \facilities{GTC (Circe), NICER, NOT (ALFOSC), NuSTAR, TNG(SiFAP), TJO (MEIA2), XMM}

\appendix

\section{SiFAP observations of the Crab PSR}
\label{app:crab}
\begin{figure}[t!]
 \includegraphics{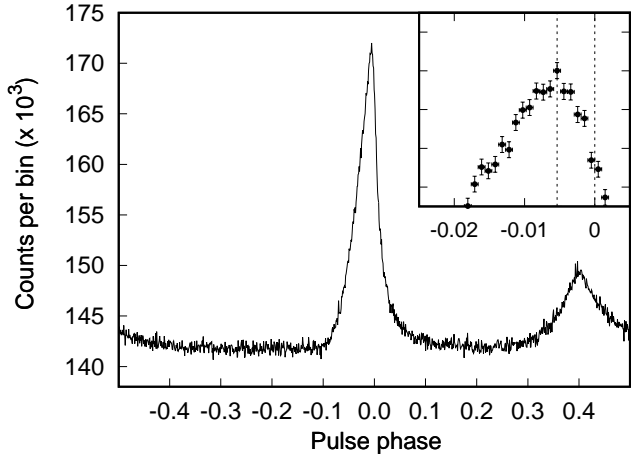}
 \caption{Pulse profile of the Crab pulsar observed with SiFAP at the
   Cassini Telescope at Loiano Observatory. Zero epoch corresponds to
   the nearest Jodrell Bank radio main pulse epoch
   $T_{JB}=57715.000000195995$~MJD. The profile was obtained folding the
   SiFAP time series in 1024 time bins using the period
   $0.03372925321514(43)$~s and the period derivative
   $4.197644(15)\times10^{-18}$ evaluated at $T_{JB}$. The inset shows
   a magnification of the peak. Vertical dotted lines indicate the
   phase of optical and radio maximum.}
 \label{fig:crab}
 \end{figure}

We observed the Crab PSR with SiFAP at the 152~cm Cassini Telescope at
Loiano Observatory for 3.3~ks starting on $T_0=$57724.04514~MJD. The
target was observed at an average count rate of
$45\times10^3$~s$^{-1}$. We reduced data following the same steps
described in Sec.~\ref{sec:tngobs}, correcting the arrival times for
the assumed linear SiFAP clock drift, and reporting them to the Solar
System barycenter using the software Tempo2 \citep{hobbs2006},
considering the position (RA $05^h$~$34^m$~$31.97232^s$ , Dec
$+22^{\circ}$~00'~$52.0690''$ [J2000]) and the JPL DE200 ephemerides
used by the Jodrell Bank monthly ephemerides
\citep{lyne1993}\footnote{Available at
  \url{http://www.jb.man.ac.uk/~pulsar/crab.html}}. We epoch folded
the optical time series in 1024 phase bins using the epoch of the
nearest arrival time of the main pulse, $T_{JB}=57715.000000195995$~MJD,
the period $0.03372925321514(43)$~s and the period derivative
$4.197644(15)\times10^{-18}$. The maximum of the main pulse occurs at
phase $\delta\phi_{Crab}=-0.0054\pm0.0005$ (see
Fig.~\ref{fig:crab}). Considering the uncertainty reported in the
Jodrell Bank ephemerides is 60$\mu$~s, we then estimate that the optical
pulse lag the radio one by $\delta\tau_{Crab}=(181\pm62)\,\mu$s. This
is compatible with the estimates given by \citet[][]{Oosterbroek2008},
who quoted $(255\pm21)\,\mu$s from simultaneous optical and radio
timing, \citet[][$\delta\tau\simeq230\,\mu$s]{germana2012}, 
\citet[][$\delta\tau\sim178\mu$s]{collins2012} and \citet[][$\delta\tau\sim240\mu$s]{zampieri2014}. We then conclude that
the absolute timing accuracy of SiFAP is better than 60$\mu$s.

\subsection{The optical flux}
\label{sec:optflux}
To determine the optical flux of {\1023} during the {\it SiFAP/TNG}
observations we first calibrated the flux/count-rate conversion factor,
$k$. The count rate observed by {\it SiFAP} is expressed by:
\begin{equation}
R=C\times\int \frac{\lambda F_{\lambda}(\lambda)}{h c} A_{e}
R(\lambda)  E(\lambda) d\lambda,
\label{eq:rate}
\end{equation}
where $F_{\lambda}(\lambda)$ is the flux density,
$A_{e}=9\times10^4$~cm$^2$ is the effective area of the {\tng} mirror,
$R(\lambda)$ is the response of the detector equipped with a white
filter (see Supplementary Fig.1 in \citealt{ambrosino2017}), and
$E(\lambda)$ is the atmospheric extinction (we considered an air-mass
of 2.0 and 1.3 for TNG1 and TNG2, respectively). We measured the
normalization factor $C$ by evaluating Eq.~\ref{eq:rate} for the
reference star; we calculated the integral for the synthetic spectrum
of a G2 V star \citep{pickles2010}\footnote{The spectrum is available
  at
  \url{http://www.eso.org/sci/facilities/paranal/decommissioned/isaac/tools/}},
normalized to give the observed flux density over the Sloan Digital
Sky Service {\it{g}} filter ($g=15.86(1)$~mag, corresponding to
$2.13\times10^{-14}$~erg~cm$^2$~nm$^{-1}$[=$1.64$~mJy]), and we
matched it to the observed, background-subtracted count rate
($R_{REF,TNG1}=19456.8$~s$^{-1}$, $R_{REF,TNG2}=30750.2$~s$^{-1}$),
obtaining $C_{TNG1}=0.434$ and $C_{TNG2}=0.686$.  We attribute the
difference between the normalization factors to the different
atmospheric conditions over the two nights. The conversion factor $k$
for {\1023} was then estimated as:
\begin{equation}
  k=\frac{F_{J1023}}{R}=\frac{F_{J1023,NOT}}{C\times\int \frac{\lambda
      f_{\lambda,NOT}(\lambda)}{h c} A_{e} R(\lambda) E(\lambda),
    d\lambda}
\end{equation}
yielding $k_{1}=5.17\times10^{-16}$~erg~cm$^{-2}$ and
$k_{2}=3.25\times10^{-16}$~erg~cm$^{-2}$ for TNG1 and TNG2,
respectively. The rightmost column of Tab.~\ref{tab:periods} lists the
320-900~nm flux measured in the different modes by scaling the net
observed count rates using these conversion factors and considering a
5\% uncertainty. The de-reddened flux in the 320--900~nm band is a
factor 1.22 larger than those values, considering the absorption
column measured by \citet[][$N_H=5.2\times10^{20}$~cm$^{-2}$]{CZ2014}
a ratio $A_V/E(B-V)=3.1$, and the resulting color excess
$E(B-V)=0.073$ \citep{predehl1995}. We checked that performing this
procedure to evaluate the ratio of the flux of {\1023} and the
reference star integrated over a B filter during the interval
simultaneous to the TJO observation performed with the same filter
(see Table~\ref{tab:log}), $F_{J1023}/F_{REF}=0.43$, is compatible
with the flux ratio 0.41(2) indicated by the magnitudes observed
($B_{PSR}=17.10(3)$~mag, $B_{REF}=16.20(1)$~mag,
$F_{J1023,TJO}/F_{REF,TJO}=0.44\pm0.02$).

\bibliography{1023}

\end{document}